\documentclass[10pt,twocolumn,twoside]{IEEEtran}
\usepackage{amsmath,amssymb}
\usepackage{amsthm}
\usepackage{cite}
\usepackage{hyperref}
\usepackage{tabularx}
\usepackage{array,booktabs}
\usepackage{tablefootnote}
\usepackage{standalone}
\usepackage{cleveref}
\usepackage{enumitem}
\usepackage{pstricks,graphicx}
\usepackage{etoolbox}
\usepackage{algpseudocode,algorithmicx,algorithm}
\usepackage{xr}
\usepackage{mdframed}

\newcommand{\mbf}[1]{\mathbf{#1}}
\newcommand{\lrf}[1]{\left\{#1\right\}}
\newcommand{\lrc}[1]{\left(#1\right)}
\newcommand{\lrs}[1]{\left[#1\right]}

\newcommand{\indicator}[1]{\mbf{1}_{\lrf{#1}}}
\newcommand{\rth}{\text{'th}}

\newcommand{\eye}{\mbf{I}}

\newtheorem{theorem}{Theorem}

\makeatletter
\g@addto@macro\th@remark{\thm@headpunct{\textnormal{:}}}
\makeatother
\theoremstyle{remark}
\newtheorem{remark}{Remark}

\newcommand{\Ugtrless}{%
	\mathrel{\kern0pt\mathop{\gtrless}\limits^{\mathcal{U}_\vIdxOne}_{\overline{\mathcal{U}}_\vIdxOne}}%
}

\newcommand{\gtr}{%
	\mathrel{\kern0pt\mathop{<}\limits^{\mathcal{U}_\vIdxOne}}%
}

\newcommand{\coherenceTime}{C}
\newcommand{\ulDuration}{C_u}
\newcommand{\dlDuration}{C_d}

\newcommand{\thresholdUlDuration}{\kappa}
\newcommand{\xTildeError}[1]{\mbf{e}_{#1}}
\newcommand{\pilotReuseRatio}{r}
\newcommand{\powerset}[1]{\mathcal{P}\lrc{#1}}
\newcommand*{\vIdxOne}{j} 
\newcommand*{\vIdxTwo}{m}
\newcommand*{\sIdxOne}{\ell}
\newcommand*{\sIdxTwo}{k}
\newcommand*{\sIdxThree}{n}
\newcommand*{\sIdxFour}{p}

\newcommand{\expectation}{\mathbb{E}}
\newcommand{\sinrUlNonIterative}{\mathrm{SINR}^{\mathrm{SP-ul}}}
\newcommand{\powerBackoff}{\omega}
\newcommand{\normChannelVector}{\bar{\mbf{h}}}
\newcommand{\normChannelPower}{\bar{\beta}}

\newcommand{\userUlRateSpNonIterative}{R^{\mathrm{SP-ul}}}
\newcommand{\cpUserSet}{\mathcal{U}_{\mathrm{TP}}}
\newcommand{\spUserSet}{\mathcal{U}_{\mathrm{SP}}}
\newcommand{\ulTxSymbol}{\mbf{s}}
\newcommand{\ulTxData}{\mbf{x}}
\newcommand{\spUlInterferencePower}{I}

\newcounter{MYtempeqncnt}
\newcommand{\cpSelectionMatrix}{\mbf{b}^{\mathrm{TP}}}
\newcommand{\spSelectionMatrix}{\mbf{b}^{\mathrm{SP}}}
\newcommand{\cpUlSinr}{\mathrm{SINR}^{\mathrm{TP-ul}}}

\newcommand{\hybridCpUlIci}{I^{\mathrm{TP-ul}}}

\newcommand{\hybridSpUlIci}{I^{\mathrm{SP-ul}}}

\newcommand{\hybridTotalIci}{I}
\newcommand{\hybridCompleteUserSet}{\mathcal{U}}
\newcommand{\cpUserSetInterim}{\cpUserSet'}
\newcommand{\spUserSetInterim}{\spUserSet'}
\newcommand{\Card}[1]{\mathrm{Card}\lrc{#1}}
\newcommand{\rxSinr}{\mathrm{SIR}^{\mathrm{Rx}}}

\newcommand{\rhoD}[2]
{
	\ifstrempty{#1}
	{
		\ifstrempty{#2}
		{
			\rho
		}
		{
			\rho_{#2}
		}
	}
	{
		\ifstrempty{#2}
		{
			\rho_{#1}
		}
		{
			\rho_{#1,#2}
		}				
	}
}

\newcommand{\rhoP}[2]
{
	\ifstrempty{#1}
	{
		\ifstrempty{#2}
		{
			\lambda
		}
		{
			\lambda_{#2}
		}
	}
	{
		\ifstrempty{#2}
		{
			\lambda_{#1}
		}
		{	
			\lambda_{#1,#2} 
		}
	}
}

\newcommand{\ulTotPower}[2]
{
	\ifstrempty{#1}
	{
		\ifstrempty{#2}
		{
			p_u
		}
		{
			\mu_{#2}
		}
	}
	{
		\mu_{#1,#2}
	}
}

\newcommand{\normRhoP}[2]
{
	\ifstrempty{#1}
	{
		\ifstrempty{#2}
		{
			\bar{\lambda}
		}
		{
			\bar{\lambda}_{#2}
		}
	}
	{
		\ifstrempty{#2}
		{
			\bar{\lambda}_{#1}
		}
		{	
			\bar{\lambda}_{#1,#2} 
		}
	}
}
\newcommand{\normRhoD}[2]
{
	\ifstrempty{#1}
	{
		\ifstrempty{#2}
		{
			\bar{\rho}
		}
		{
			\bar{\rho}_{#2}
		}
	}
	{
		\ifstrempty{#2}
		{
			\bar{\rho}_{#1}
		}
		{
			\bar{\rho}_{#1,#2}
		}				
	}
}

\newcommand{\spPilot}[2]
{
	\ifstrempty{#1}
	{
		\ifstrempty{#2}
		{
			\mbf{p}
		}
		{
			\mbf{p}_{#2}
		}
	}
	{
		\ifstrempty{#2}
		{
			\mbf{p}_{#1}
		}
		{
			\mbf{p}_{#1,#2}
		}				
	}
}

\title{Superimposed Pilots are Superior for Mitigating Pilot Contamination in Massive MIMO}
	\author{Karthik Upadhya,~\IEEEmembership{Student~Member,~IEEE}, 
		Sergiy A. Vorobyov,~\IEEEmembership{Senior~Member,~IEEE}, \\
		Mikko Vehkapera,~\IEEEmembership{Member,~IEEE}
		\thanks{Manuscript received June 10, 2016; revised Dec. 5, 2016; accepted Feb. 8, 2017. The associate editor coordinating the review of this manuscript and approving it for publication was Dr. Marios Kountouris.}
		\thanks{Copyright (c) 2017 IEEE. Personal use of this material is permitted. However, permission to use this material for any other purposes must be obtained from the IEEE by sending a request to pubs-permissions@ieee.org.}
		\thanks{Part of this work has been presented at the 2016 IEEE International Conf. Acoustics, Speech, and Signal Processing, Shanghai, China.}
		\thanks{This work was supported in part by the Academy of Finland research grant No. 299243.}
		\thanks{K. Upadhya and S. A. Vorobyov are with the Department of Signal Processing and Acoustics, Aalto University, FI-00076 Aalto, Finland (E-mails: karthik.upadhya@aalto.fi, svor@ieee.org).}
		\thanks{M. Vehkapera is with the Department of Electronic and Electrical Engineering, University of Sheffield, Sheffield, S1 3JD, UK (E-mail: m.vehkapera@sheffield.ac.uk).}
	}

\begin{document}
\maketitle
\begin{abstract}
	In this paper, superimposed pilots are introduced as an alternative to time-multiplexed pilot and data symbols for mitigating pilot contamination in massive multiple-input multiple-output (MIMO) systems. We propose a non-iterative scheme for uplink channel estimation based on superimposed pilots and derive an expression for the uplink signal-to-interference-plus-noise ratio (SINR) at the output of a matched filter employing this channel estimate. Based on this expression, we observe that power control is essential when superimposed pilots are employed. Moreover, the quality of the channel estimate can be improved by reducing the interference that results from transmitting data alongside the pilots, and an intuitive iterative data-aided scheme that reduces this component of interference is also proposed. Approximate expressions for the uplink SINR are provided for the iterative data-aided method as well. In addition, we show that a hybrid system with users utilizing both time-multiplexed and superimposed pilots is superior to an optimally designed system that employs only time-multiplexed pilots, even when the non-iterative channel estimate is used to build the detector and precoder. We also describe a simple approach to implement this hybrid system by minimizing the overall inter and intra-cell interference. Numerical simulations demonstrating the performance of the proposed channel estimation schemes and the superiority of the hybrid system are also provided. 
\end{abstract}



\section{Introduction}
Massive multiple-input multiple-output (MIMO) systems, proposed in \cite{Marzetta2010Noncooperative}, have received significant interest in recent years as a candidate for fifth-generation mobile communication technologies \cite{boccardi2014fivedisruptive,larsson2014massive,jeffandrews2014whatwill}. These systems are a variation of multi-user MIMO (MU-MIMO) and have a large number of base station (BS) antennas that result in an improved spectral efficiency through spatial multiplexing \cite{lulu2014anoverview,rusek2013scaling}. Under favorable propagation conditions \cite{Marzetta2010Noncooperative}, significant gains in throughput can be achieved by employing simple linear processing at the BS \cite{hoydis2013massive,yang2013performance}. In addition, large numbers of antennas result in an improved uplink (UL) energy-efficiency \cite{ngo2013energy} and render the system performance resilient to hardware impairments \cite{bjornson2014massive}.

However, all the above mentioned benefits of a massive MIMO communication system hinge on the assumption that the BS has access to accurate estimates of the channel state information (CSI). For systems that employ either frequency division duplexing (FDD) or time division duplexing (TDD), the channel estimates are obtained using orthogonal pilot sequences. In FDD systems, each antenna at the BS transmits a pilot sequence that is orthogonal to the pilot sequences transmitted by the other antennas. Since the number of orthogonal pilot sequences required becomes proportional to the number of BS antennas, FDD is considered impractical for channel estimation in massive MIMO \cite{lulu2014anoverview,bjornson2016massive}. Moreover, since the CSI corresponding to each antenna is estimated by the users, it has to be fedback from the users to the BS, consuming additional bandwidth. Consequently, massive MIMO systems are typically assumed to employ TDD with full frequency-reuse and utilize channel reciprocity to obtain CSI. In these systems, each user in a cell is assigned a different pilot sequence and these pilots are time-multiplexed with data in each coherence block. When using time-multiplexed pilots and data, the requirement for high spectral efficiency necessitates sharing of pilot sequences between adjacent cells, resulting in the channel estimates of the users in a cell being corrupted by the channel vectors of users in the adjacent cells. This phenomenon called `pilot contamination' \cite{jose2011pilot} introduces interference in both the UL and downlink (DL), degrading the overall performance of the system.

Existing methods to mitigate pilot contamination for massive MIMO are designed for the case wherein the pilots are time-multiplexed with the data, henceforth referred to as time-multiplexed pilots. In \cite{ngo2012evd}, it has been observed that the eigenvectors of the autocorrelation matrix of the received data correspond to the channel vectors of the desired and interfering users, and a method for channel estimation has been developed based on this observation. Pilot decontamination has been performed in \cite{Muller2014Blind} by projecting the contaminated channel estimate on an interference-free subspace spanned by the channel vectors of the desired users, whereas \cite{vinogradova2016separability} derives asymptotic conditions for separability between the subspaces of the desired and interfering users. In \cite{Yin2013Coordinated}, a coordinated method for pilot allocation has been proposed for separating desired and interfering users in correlated channels. A pilot decontamination method based on the array processing model has been proposed in \cite{upadhya2015anarray} for use in parametric channels with a finite number of discrete paths. In \cite{bjornson2015massive}, a resource allocation approach has been proposed for optimizing the number of users scheduled in each cell in order to minimize the effect of pilot contamination. 
A common theme for the approaches described above, except for \cite{Yin2013Coordinated} and \cite{bjornson2015massive}, is that they focus on decontaminating the channel estimate at the receiver, which in this case is the BS. However, since pilot contamination results from interfering pilot transmissions, there is a scope for \emph{better separating the desired and interfering users by optimizing the pilot transmissions at the user terminal} as well. 

In this paper, we propose using superimposed pilots as an alternative to, and in combination with, time-multiplexed pilots in massive MIMO systems. Methods for channel estimation based on pilots that are embedded in data, such as superimposed pilots, have been extensively studied for MIMO systems \cite{takeuchi2013Achievable,shuangchi2007superimposed,coldrey2007training,haidong2003pilot,cui2005pilot,ghogho2005channel}. However, these papers have focused on embedded and superimposed pilots in the context of accommodating a loss in signal-to-noise ratio (SNR) in exchange for a reduced pilot transmission overhead \cite{shuangchi2007superimposed,coldrey2007training}. Particularly, scenarios with high user-mobility, wherein it is impractical to allocate dedicated symbols for training, have been of interest for employing superimposed pilots.
In the context of multi-cell massive MIMO, provided that the number of users in the system is smaller than the number of symbols in the UL, superimposed pilots allow for each user in the system to be assigned a unique pilot sequence, enabling the receiver to estimate the channel vectors of both the desired and interfering users. In addition, these pilots mitigate pilot contamination by time-averaging over long sequences and offer a higher efficiency due to a reduced transmission overhead. 

We obtain expressions for the signal-to-interference-plus-noise ratio (SINR) at the output of a matched filter (MF)-based detector when a non-iterative least-squares (LS)-based channel estimate is employed for channel estimation. Based on the SINR expression, we highlight the need for power control when superimposed pilots are employed in a massive MIMO system. Moreover, we discuss the shortcomings of the non-iterative channel estimator and propose an intuitive low-complexity iterative channel estimation scheme for superimposed pilots.\footnote{The work in this paper is a significant extension of our relevant conference paper \cite{upadhya2015superimposed}. In addition to a detailed exposition, we have included additional results that demonstrate the superiority of massive MIMO systems that use superimposed pilots instead of time-multiplexed pilots.
} In addition, we introduce the concept of a hybrid system and show by means of theoretical arguments that the hybrid system is superior to its counterpart that employs only time-multiplexed pilots, even when the non-iterative channel estimator is used to obtain the channel estimate from superimposed pilots. A simple approach to design and implement this hybrid system is also detailed. Although the use of superimposed pilots requires some coordination between the BSs in assigning pilot sequences to the users and estimating their path-loss coefficients, these are minor impediments compared to the performance improvements provided by the proposed scheme.

The article in the existing literature that is closest to this paper is \cite{Zhang2015SuperimposedPilot}, wherein superimposed pilots have been employed in the context of multi-cell multiuser MIMO systems. However, unlike \cite{Zhang2015SuperimposedPilot}, the focus of our paper is to demonstrate the superiority of superimposed pilots when used in conjunction with time-multiplexed pilots in a hybrid system. The theoretical results and simulations that have been obtained are in line with this objective. 

In Section \ref{sec:systemModel}, the system model for the massive MIMO UL is described. In Section \ref{sec:conventionalPilots}, time-multiplexed pilots are described and the pilot contamination problem is detailed. \mbox{Section \ref{sec:superimposedPilots}} introduces the superimposed pilot scheme and describes the non-iterative method for channel estimation and Section \ref{sec:superimposedPilotsIterative} discusses the iterative data-aided scheme. 
In Section \ref{sec:hybridSystem}, the concept of a hybrid system that employs both time-multiplexed and superimposed pilots is introduced and in Section \ref{sec:hybridSystemImplementation}, a simple approach for implementing this hybrid system is discussed.
Section \ref{sec:numericalSimulations} presents simulation results demonstrating the effectiveness of employing superimposed pilots for pilot decontamination. 
Section \ref{sec:conclusion} concludes the paper. Some of the proofs are given in Appendix.

\textit{Notation} : Lower case and upper case boldface letters denote column vectors and matrices, respectively. The notations $(\cdot)^*,(\cdot)^T$, $ (\cdot)^H $, and $ (\cdot)^{-1} $ represent the conjugate, transpose, Hermitian transpose, and inverse, respectively. The notation $ \mathcal{C}\mathcal{N}(\pmb{\mu},\mbf{\Sigma}) $ stands for the complex normal distribution with mean $ \pmb{\mu} $ and covariance matrix $ \mbf{\Sigma} $ and $ \mathbb{E}\lrf{\cdot} $ is used to denote the expectation operator. The notation $ \eye_{N} $ denotes an $ N\times N $ identity matrix, and $ \|\cdot\| $ and $ \|\cdot\|_F  $ denote the Euclidean norm of a vector and Frobenius norm of a matrix, respectively. Upper case calligraphic letters denote sets, and $ \varnothing $ denotes the empty set. The notation $ \indicator{\mathcal{S}} $ represents the indicator function 	over the set $ \mathcal{S} $, whereas $ \Card{\mathcal{S}} $ is used to represent its cardinality. 
The notation $ \delta_{n,m} $ denotes the Kronecker delta function, and $ \eta(\cdot) $ stands for an element-by-element decision function that replaces each element of the input vector with the constellation point that is closest in Euclidean distance to that element. 
The big O notation $ f(x) = O(g(x)) $ implies that $ |f(x)|/|g(x)| $ is bounded as $ x\rightarrow\infty $.


\section{System Model}
\renewcommand{\vIdxOne}{j}
\renewcommand{\vIdxTwo}{m}
\renewcommand{\sIdxOne}{\ell}
\renewcommand{\sIdxTwo}{k}
\label{sec:systemModel}
We consider a TDD massive MIMO UL with $ L $ cells and $ K $ single-antenna\footnote{ For training and channel estimation, users with $ T>1 $ antennas can be treated as $ T $ separate single-antenna users.} users per cell. Each cell has a BS with $ M \gg K $ antennas. The number of symbols $ \coherenceTime $, over which the channel is coherent, is assumed to be divided into $ \ulDuration $ and $ \dlDuration $, which are the number of symbols in the UL and DL time slots, respectively.
The matrix of received measurements $ \mbf{Y}_{\vIdxOne}\in\mathbb{C}^{M\times \ulDuration} $ at BS $ \vIdxOne $ can be written as
\begin{equation}
\mbf{Y}_{\vIdxOne} = \sum\limits_{\sIdxOne=0}^{L-1} \sum\limits_{\sIdxTwo=0}^{K-1} \sqrt{\ulTotPower{\sIdxOne}{\sIdxTwo}} \mbf{h}_{\vIdxOne,\sIdxOne,\sIdxTwo} \ulTxSymbol_{\sIdxOne,\sIdxTwo}^T + \mbf{W}_{\vIdxOne}
\end{equation}
where $ \ulTotPower{\sIdxOne}{\sIdxTwo} $ denotes the power with which the vector of symbols  \mbox{$ \ulTxSymbol_{\sIdxOne,\sIdxTwo}\in\mathbb{C}^{C_u\times 1} $} are transmitted by user~$ \sIdxTwo $ in cell~$ \sIdxOne $, $ \mbf{W}_\vIdxOne\in\mathbb{C}^{M\times C_u} $ is the additive white Gaussian noise at BS $ \vIdxOne $ with each column distributed as $ \mathcal{CN}(\mbf{0},\sigma^2\eye_{M}) $. Moreover, the columns of $ \mbf{W}_\vIdxOne $ are mutually independent of each other. 
The vector $ \mbf{h}_{\vIdxOne,\sIdxOne,\sIdxTwo} \in \mathbb{C}^{M\times 1} $ represents the channel response between the antennas at BS $ \vIdxOne $, and user $ \sIdxTwo $ in cell $ \sIdxOne $, and is assumed to be distributed as\footnote{While, for the sake of simplicity, an environment with rich scattering is assumed, the conclusions made in this paper are independent of the channel distribution and only require the channel vectors of any pair of users to be asymptotically orthogonal.} 
\begin{equation}
\mbf{h}_{\vIdxOne,\sIdxOne,\sIdxTwo} \sim \mathcal{CN}(\mbf{0},\beta_{\vIdxOne,\sIdxOne,\sIdxTwo}\eye_{M})
\end{equation}
where $ \beta_{\vIdxOne,\sIdxOne,\sIdxTwo} $ denotes the large-scale path-loss coefficient which depends on the user location in the cell. In addition, the channel vectors $ \mbf{h}_{\vIdxOne,\sIdxOne,\sIdxTwo} $ are assumed to be mutually independent of each other $ \forall \vIdxOne,\sIdxOne,\sIdxTwo $. The aforementioned statistics of the channel vector correspond to the non-line-of-sight scenario with rich scattering \cite{Marzetta2010Noncooperative}. By virtue of their zero mean and mutual independence, the channel vectors are asymptotically orthogonal and the following equation holds almost surely \cite{Marzetta2010Noncooperative}
\begin{equation}
\lim\limits_{M\rightarrow\infty}\frac{\mbf{h}_{j,\ell,k}^H \mbf{h}_{m,n,p}}{M}=\ \beta_{j,\ell,k}\;\delta_{j,m} \delta_{\ell,n} \delta_{k,p}, \;\forall\;j,k,\ell,m,n,p \;.
\end{equation} 
Moreover, $ \mbf{h}_{\vIdxOne,\sIdxOne,\sIdxTwo} $ is assumed to be constant for the duration of $ \coherenceTime $ symbols, and $ \beta_{\vIdxOne,\sIdxOne,\sIdxTwo} $ is constant for a significantly longer duration which depends on the user mobility. For the sake of simplicity, the effects of shadowing are not taken into account in this paper.\footnote{The algorithms and analysis in this paper remain the same in the presence of shadowing, provided that the users are allocated to the strongest BSs. However, the geometric interpretations that are made based on the location of the user in the cell will no longer be valid.}  The transmitted symbols $ \ulTxSymbol_{\sIdxOne,\sIdxTwo} $ contain both pilots and data. The pilots could either be time-multiplexed or superimposed pilots, and the elements of the data vector $ \mbf{x}_{\sIdxOne,\sIdxTwo} $ are assumed to be independent and identically distributed (i.i.d) random variables with zero-mean and unit variance and take values from an alphabet $ \chi $, which is a realistic assumption.

\section{Time-Multiplexed Pilots And The Pilot Contamination Problem}
\label{sec:conventionalPilots}
With time-multiplexed pilots, each user in a cell transmits a $ \tau \geq K $ length orthogonal pilot sequence for channel estimation followed by $ \ulDuration-\tau $ symbols of uplink data. In order to minimize the overhead incurred, it is necessary to reuse these pilot sequences in the adjacent cells. However, this pilot-reuse results in the channel estimates of the desired users being contaminated by the channel vectors of users in adjacent cells, causing interference and in turn, a loss in spectral efficiency. 

It is assumed here that the transmission of the pilot sequences by the users in the $ L $ cells are synchronized, which corresponds to the worst-case scenario for pilot contamination.\footnote{No additional improvement in the UL performance can be gleaned by separating the pilot and data transmissions across cells \cite{Marzetta2010Noncooperative,ngo2013energy,bjornson2015massive}.}
Consider a unitary matrix $ \mbf{\Phi} \in \mathbb{C}^{\tau\times\tau} $ whose columns $ \lrf{\pmb{\phi}_1,\ldots,\pmb{\phi}_\tau} $ are the orthogonal pilot sequences that are transmitted by the users, i.e., $ \pmb{\phi}_{\sIdxThree}^H\pmb{\phi}_{\sIdxFour} = \tau\delta_{\sIdxThree,\sIdxFour} $. 
If $ \pmb{\phi}_{b_{\sIdxOne,\sIdxTwo}} $ is the pilot sequence transmitted by user~$ \sIdxTwo $ of cell~$ \sIdxOne $, where \mbox{$ b_{\sIdxOne,\sIdxTwo} \in \lrf{1,\ldots,\tau} $} is the index of the transmitted pilot,
and if each pilot sequence is reused once every $ \pilotReuseRatio\triangleq \tau/K $ cells \cite{bjornson2015massive}, the LS estimate of the channel of user $ \vIdxTwo $ in cell $ \vIdxOne $ can be obtained as \cite{Marzetta2010Noncooperative}
\begin{equation}
\label{eqn:pilotContamination}
\widehat{\mbf{h}}_{\vIdxOne,\vIdxOne,\vIdxTwo}^{\mathrm{TP}} \!\! \triangleq \! \frac{1}{\tau\sqrt{\ulTotPower{}{}}} \mbf{Y}_\vIdxOne^{(p)}\pmb{\phi}_{b_{\vIdxOne,\vIdxTwo}}^*
\!\!\!\!\! = \!
\mbf{h}_{\vIdxOne,\vIdxOne,\vIdxTwo}
+
\!\!\!\!
\sum\limits_{\substack{\sIdxOne=0\\\sIdxOne\neq \vIdxOne\\\sIdxOne\in\mathcal{L}_\vIdxOne(\pilotReuseRatio)}}^{L-1} 
\!\!\!
\mbf{h}_{\vIdxOne,\sIdxOne,\vIdxTwo} 
+ 
\frac{1}{\tau\sqrt{\ulTotPower{}{}}}\!\mbf{W}_{\!\vIdxOne}^{(p)}\pmb{\phi}_{b_{\vIdxOne,\vIdxTwo}}^*
\end{equation}
where the superscript TP indicates that the estimates are computed when using time-multiplexed pilots, the superscript $ p $ indicates that the observations are made during pilot transmission, and $ \mathcal{L}_\vIdxOne(\pilotReuseRatio) $ is the subset of the $ L $ cells that use the same set of pilots as cell $ \vIdxOne $. In addition, it is assumed in $ \eqref{eqn:pilotContamination} $ without loss of generality that the transmit powers are same for all users employing time-multiplexed pilots, i.e., \mbox{$ \ulTotPower{\sIdxOne}{\sIdxTwo} = \ulTotPower{}{},\;\forall\sIdxOne,\sIdxTwo $}, and any variation in the transmit power of an individual user is absorbed into the corresponding path-loss coefficient $ \beta $. It can be observed from \eqref{eqn:pilotContamination} that the estimates of the channel vectors of the users in cell $ \vIdxOne $ are contaminated by the channel vectors of the users in the remaining $ \Card{\mathcal{L}_{\vIdxOne}(r) - 1} $ cells. 
When \mbox{$ M\rightarrow\infty $}, the UL SINR of user $ \vIdxTwo $ in cell $ \vIdxOne $, at the output of an MF that uses the channel estimate in \eqref{eqn:pilotContamination} for detection, can be written as \cite{Marzetta2010Noncooperative}
\begin{align}
\textrm{SINR}_{\vIdxOne,\vIdxTwo}^{\mathrm{TP-ul}} =  \frac{\beta^2_{\vIdxOne,\vIdxOne,\vIdxTwo}}{\sum\limits_{\substack{\sIdxOne\neq \vIdxOne\\\sIdxOne\in\mathcal{L}_\vIdxOne(\pilotReuseRatio)}} \beta^2_{\vIdxOne,\sIdxOne,\vIdxTwo} }\;. \label{eqn:sinrConventionalPilot}
\end{align}
The corresponding throughput of the user using Gaussian signaling in the UL can then be expressed as \cite{Marzetta2010Noncooperative}
\begin{align}
\textrm{R}_{\vIdxOne,\vIdxTwo}^{\mathrm{TP-ul}} &= \frac{(\ulDuration-\tau)}{\coherenceTime} \log_2\lrc{1+\textrm{SINR}_{\vIdxOne,\vIdxTwo}^{\mathrm{TP-ul}}} 
\;. \label{eqn:rateConventionalPilots}
\end{align}
From the above equation, it can be observed that the rate per user is a function of both the  overhead $ \tau $ as well as the loss in SINR due to pilot contamination. A larger value of $ \pilotReuseRatio $ would reduce the effect of pilot contamination and increase the SINR at the cost of a reduced transmission efficiency $ (\ulDuration-\tau)/\coherenceTime $. 


\section{Superimposed Pilots}

\begin{figure*}
	\normalsize
	\setcounter{MYtempeqncnt}{\value{equation}}
	\setcounter{equation}{11}
	
	\begin{align}
	\sinrUlNonIterative_{\vIdxOne,\vIdxTwo}
	&=
	\lrc
	{
		\sum\limits_{\sIdxOne=0}^{L-1}
		\sum\limits_{\sIdxTwo=0}^{K-1}
		\frac{
			\rhoD{\sIdxOne}{\sIdxTwo}^2
			\ulTotPower{\sIdxOne}{\sIdxTwo}	
			\beta_{\vIdxOne,\sIdxOne,\sIdxTwo}^2
		}
		{
			\ulDuration
			\rhoP{\vIdxOne}{\vIdxTwo}^2
			\rhoD{\vIdxOne}{\vIdxTwo}^2
			\beta_{\vIdxOne,\vIdxOne,\vIdxTwo}^2
		}
		+
		\frac{1}{M}
		\lrc{
			\mathop
			{
				\sum\limits_{\sIdxOne=0}^{L-1}
				\sum\limits_{\sIdxTwo=0}^{K-1}
			}\limits_{\lrf{\sIdxOne\neq\vIdxOne,\sIdxTwo\neq\vIdxTwo}}
			\frac{
				\beta_{\vIdxOne,\sIdxOne,\sIdxTwo}
				\ulTotPower{\sIdxOne}{\sIdxTwo}
			}
			{
				\rhoD{\vIdxOne}{\vIdxTwo}^2
				\beta_{\vIdxOne,\vIdxOne,\vIdxTwo}
			}
			+
			\mathop
			{
				\sum\limits_{\sIdxOne=0}^{L-1}
				\sum\limits_{\sIdxTwo=0}^{K-1}
			}\limits_{\lrf{\sIdxOne\neq\vIdxOne,\sIdxTwo\neq\vIdxTwo}}
			\mathop
			{
				\sum\limits_{\sIdxThree=0}^{L-1}
				\sum\limits_{\sIdxFour=0}^{K-1}
			}\limits_{\lrf{\sIdxThree\neq\sIdxOne,\sIdxFour\neq\sIdxTwo}}
			\frac{
				\rhoD{\sIdxThree}{\sIdxFour}^2
				\beta_{\vIdxOne,\sIdxOne,\sIdxTwo}
				\beta_{\vIdxOne,\sIdxThree,\sIdxFour}
				\ulTotPower{\sIdxOne}{\sIdxTwo}
			}
			{
				\ulDuration
				\rhoP{\vIdxOne}{\vIdxTwo}^2
				\rhoD{\vIdxOne}{\vIdxTwo}^2
				\beta_{\vIdxOne,\vIdxOne,\vIdxTwo}^2
			}
		}
	}^{-1}
	\label{eqn:sinrNonIterativeFiniteM}
	\end{align}
	\setcounter{equation}{\value{MYtempeqncnt}}
	\hrulefill
	\vspace*{-10pt}
\end{figure*}

\label{sec:superimposedPilots}
With superimposed pilots, the pilot symbols are transmitted at a reduced power alongside the data symbols, and in its simplest version, the pilot and data symbols are transmitted alongside each other for the entire duration of the uplink data slot $ \ulDuration $. If the total number of users in the system is smaller than the number of symbols in the uplink, i.e., $ KL \leq \ulDuration $,\footnote{ For example, using the orthogonal frequency division multiplexing (OFDM) parameters in long-term evolution (LTE) systems as in \cite{Marzetta2010Noncooperative}, i.e., $ \ulDuration = 7 $ OFDM symbols, $ N_{\mathrm{smooth}} = 14 $ subcarriers, and assuming the pilots are reused over $ {L=7} $ hexagonal cells, the maximum number of supported users in the $ L $ cells is $ \ulDuration N_{\mathrm{smooth}} = 98 $ users. Therefore, the number of users per cell is $ \ulDuration N_{\mathrm{smooth}}/L = 14 $ users. However, note that the value of $ \ulDuration =7 $ has been chosen to allow user velocities of 350 km/h \cite{cox2012introduction}. For lower user speeds and with cell sectoring, larger number users can be supported and the assumption $ KL\leq \ulDuration $ will easily be satisfied.} then with superimposed pilots, each user can be assigned a unique orthogonal pilot $ \spPilot{\sIdxOne}{\sIdxTwo} \in \mathbb{C}^{\ulDuration\times 1} $. The pilots are taken from the columns of a unitary matrix $ \mbf{P} \in \mathbb{C}^{\ulDuration\times\ulDuration} $ such that $ \mbf{P}^H\mbf{P} = \ulDuration \eye_{\ulDuration} $, and therefore $ \spPilot{\sIdxOne}{\sIdxTwo}^H\spPilot{\sIdxThree}{\sIdxFour} = \ulDuration \delta_{\sIdxOne,\sIdxThree} \delta_{\sIdxTwo,\sIdxFour} $. If $ {\rhoD{\sIdxOne}{\sIdxTwo}\mbf{x}_{\sIdxOne,\sIdxTwo}+\rhoP{\sIdxOne}{\sIdxTwo}\mbf{p}_{\sIdxOne,\sIdxTwo}} $ is the transmitted vector from user $ \sIdxTwo $ in cell $ \sIdxOne $, then
the received signal at the $ \vIdxOne\rth $ BS $ \mbf{Y}_{\vIdxOne}\in\mathbb{C}^{M\times \ulDuration} $, when using the superimposed pilot scheme, can be written as
\begin{equation}
\label{eqn:superimposedPilotsDefn}
\mbf{Y}_{\vIdxOne} 
= 
\sum\limits_{\sIdxOne=0}^{L-1}
\sum\limits_{\sIdxTwo=0}^{K-1} 
\mbf{h}_{\vIdxOne,\sIdxOne,\sIdxTwo} 
\lrc{
	\rhoD{\sIdxOne}{\sIdxTwo}
	\mbf{x}_{\sIdxOne,\sIdxTwo}
	+
	\rhoP{\sIdxOne}{\sIdxTwo}
	\spPilot{\sIdxOne}{\sIdxTwo}
}^T 
+ 
\mbf{W}_{\vIdxOne}
\end{equation}
where $ \rhoP{\sIdxOne}{\sIdxTwo}^2$ and $\rhoD{\sIdxOne}{\sIdxTwo}^2$ are the fractions of the transmit power reserved for the pilot and data symbols, respectively, and the total transmitted power $ \ulTotPower{\sIdxOne}{\sIdxTwo} $ is given as $ \ulTotPower{\sIdxOne}{\sIdxTwo} = \rhoP{\sIdxOne}{\sIdxTwo}^2+\rhoD{\sIdxOne}{\sIdxTwo}^2 $.
\subsection{Non-Iterative Channel Estimation}
\label{subsec:nonIterativeChannelEstimation}
Treating the data symbols of all users as additive noise, the channel estimate of user $ \sIdxTwo $ in cell $ \sIdxOne $ can be obtained at the $ \vIdxOne\rth $ BS using the LS criterion \cite{Zhang2015SuperimposedPilot}
\renewcommand{\sIdxThree}{m}
\renewcommand{\sIdxFour}{n}
\begin{align}
\label{eqn:superimposedNonIterativeLSCriterion}
\widehat{\mbf{h}}_{\vIdxOne,\sIdxOne,\sIdxTwo} 
&\triangleq 
\arg\underset{\mbf{h}}{\min}{\|\mbf{Y}_\vIdxOne - \rhoP{\sIdxOne}{\sIdxTwo} \mbf{h}\; \spPilot{\sIdxOne}{\sIdxTwo}^T \|^2_F} \;.
\end{align}
Solving \eqref{eqn:superimposedNonIterativeLSCriterion} yields
\begin{align}
\label{eqn:superimposedPilotNonIterativeChannel}
\widehat{\mbf{h}}_{\vIdxOne,\sIdxOne,\sIdxTwo}
&=
\mbf{Y}_{\vIdxOne}
\lrc{
	\rhoP{\sIdxOne}{\sIdxTwo}^2
	\;
	\spPilot{\sIdxOne}{\sIdxTwo}^H
	\spPilot{\sIdxOne}{\sIdxTwo}
}^{-1}
\rhoP{\sIdxOne}{\sIdxTwo} 
\spPilot{\sIdxOne}{\sIdxTwo}^*
= 
\frac{1}{\ulDuration\rhoP{\sIdxOne}{\sIdxTwo}}
\mbf{Y}_{\vIdxOne} 
\spPilot{\sIdxOne}{\sIdxTwo}^* 
\nonumber
\\
&= 
\mbf{h}_{\vIdxOne,\sIdxOne,\sIdxTwo}
+ 
\frac{1}{\ulDuration\rhoP{\sIdxOne}{\sIdxTwo}} 
\sum\limits_{\sIdxThree=0}^{L-1} 
\sum\limits_{\sIdxFour=0}^{K-1} 
\rhoD{\sIdxThree}{\sIdxFour}
\mbf{h}_{\vIdxOne,\sIdxThree,\sIdxFour} 
\mbf{x}_{\sIdxThree,\sIdxFour}^T
\spPilot{\sIdxOne}{\sIdxTwo}^*  
\nonumber 
\\
&\quad
+ 
\frac{1}{\ulDuration\rhoP{\sIdxOne}{\sIdxTwo}} 
\mbf{W}_{\vIdxOne}
\spPilot{\sIdxOne}{\sIdxTwo}^* \;.
\end{align}
In order to estimate the data from the received observations, it is necessary to remove the term corresponding to the transmitted superimposed pilot $ \rhoP{\vIdxOne}{\vIdxTwo} \mbf{h}_{\vIdxOne,\vIdxOne,\vIdxTwo} \spPilot{\vIdxOne}{\vIdxTwo}^T $ from the observation vector in \eqref{eqn:superimposedPilotsDefn}. Using $ \rhoP{\vIdxOne}{\vIdxTwo} \widehat{\mbf{h}}_{\vIdxOne,\vIdxOne,\vIdxTwo} \spPilot{\vIdxOne}{\vIdxTwo}^T $ as an estimate for this term, the estimate of $ \mbf{x}_{\vIdxOne,\vIdxTwo} $ can then be obtained from the observation $ \mbf{Y}_\vIdxOne $ using an MF and a decision operation as follows 
\begin{align}
\label{eqn:superimposedPilotNonIterativeMF}
\widetilde{\mbf{x}}_{\vIdxOne,\vIdxOne,\vIdxTwo}^T 	 
&= 
{
	\frac{1}{M\rhoD{\vIdxOne}{\vIdxTwo}
		\beta_{\vIdxOne,\vIdxOne,\vIdxTwo}} 
	\widehat{\mbf{h}}_{\vIdxOne,\vIdxOne,\vIdxTwo}^H 
	\lrc{
		\mbf{Y}_{\vIdxOne} 
		- 
		\rhoP{\vIdxOne}{\vIdxTwo} 
		\widehat{\mbf{h}}_{\vIdxOne,\vIdxOne,\vIdxTwo} 
		\mbf{p}_{\vIdxOne,\vIdxTwo}^T}
}  \\
\widehat{\mbf{x}}_{\vIdxOne,\vIdxOne,\vIdxTwo} 	 
&= 
\eta\lrc{\widetilde{\mbf{x}}_{\vIdxOne,\vIdxOne,\vIdxTwo}}
\label{eqn:superimposedPilotNonIterativeHardSlicing}\;.
\end{align}
The SINR of user $ \vIdxTwo $ in cell $ \vIdxOne $, at the output of an MF that employs the channel estimate in \eqref{eqn:superimposedPilotNonIterativeChannel}, is derived in Appendix~\ref{appdx:sinrNonIterative} and is given in \eqref{eqn:sinrNonIterativeFiniteM} (shown at the top of the page). The SINR in \eqref{eqn:sinrNonIterativeFiniteM}, when $ M\rightarrow\infty $, can be written as
\setcounter{equation}{\value{equation}+1}
\begin{align}
\sinrUlNonIterative_{\vIdxOne,\vIdxTwo}
= 
\frac{\rhoP{\vIdxOne}{\vIdxTwo}^2\rhoD{\vIdxOne}{\vIdxTwo}^2\beta_{\vIdxOne,\vIdxOne,\vIdxTwo}^2}
{
	\frac{1}{\ulDuration}
	\sum\limits_{\sIdxOne=0}^{L-1}
	\sum\limits_{\sIdxTwo=0}^{K-1}
	\rhoD{\sIdxOne}{\sIdxTwo}^2
	\ulTotPower{\sIdxOne}{\sIdxTwo}
	\beta_{\vIdxOne,\sIdxOne,\sIdxTwo}^2
}
\ .\label{eqn:sinrSuperimposedPilotNonIterativeInfM}
\end{align}
The corresponding per-user rate in the uplink when using Gaussian signaling is given as
\begin{align}
\userUlRateSpNonIterative_{\vIdxOne,\vIdxTwo} &=
\frac{\ulDuration}{\coherenceTime} \log_2\lrc{1+\textrm{SINR}_{\vIdxOne,\vIdxTwo}^{\mathrm{SP-ul}}} \;. 
\label{eqn:rateSuperimposedPilotNonIterative}
\end{align}

\begin{figure*}
	\normalsize
	\setcounter{MYtempeqncnt}{\value{equation}}
	\setcounter{equation}{21}
	
	\begin{align}
	\sinrUlNonIterative_{\vIdxOne,\vIdxTwo}
	\!\!
	&=
	\!\!
	\lrc
	{	
		\!
		\sum\limits_{\sIdxOne=0}^{L-1}
		\!
		\sum\limits_{\sIdxTwo=0}^{K-1}
		\!
		\frac{
			\normRhoD{\sIdxOne}{\sIdxTwo}^2
			\normChannelPower_{\vIdxOne,\sIdxOne,\sIdxTwo}^2
		}
		{
			\ulDuration
			\normRhoP{\vIdxOne}{\vIdxTwo}^2
			\normRhoD{\vIdxOne}{\vIdxTwo}^2
			\normChannelPower_{\vIdxOne,\vIdxOne,\vIdxTwo}^2
		}
		\!
		+
		\!
		\frac{1}{M}
		\!
		\lrc{
			\mathop
			{
				\sum\limits_{\sIdxOne=0}^{L-1}
				\sum\limits_{\sIdxTwo=0}^{K-1}
			}\limits_{\lrf{\sIdxOne\neq\vIdxOne,\sIdxTwo\neq\vIdxTwo}}
			\frac{
				\normChannelPower_{\vIdxOne,\sIdxOne,\sIdxTwo}
			}
			{
				\normRhoD{\vIdxOne}{\vIdxTwo}^2
				\normChannelPower_{\vIdxOne,\vIdxOne,\vIdxTwo}
			}
			\!
			+
			\!
			\mathop
			{
				\sum\limits_{\sIdxOne=0}^{L-1}
				\!
				\sum\limits_{\sIdxTwo=0}^{K-1}
			}\limits_{\lrf{\sIdxOne\neq\vIdxOne,\sIdxTwo\neq\vIdxTwo}}
			\mathop
			{
				\sum\limits_{\sIdxThree=0}^{L-1}
				\sum\limits_{\sIdxFour=0}^{K-1}
			}\limits_{\lrf{\sIdxThree\neq\sIdxOne,\sIdxFour\neq\sIdxTwo}}
			\frac{
				\normRhoD{\sIdxThree}{\sIdxFour}^2
				\normChannelPower_{\vIdxOne,\sIdxOne,\sIdxTwo}
				\normChannelPower_{\vIdxOne,\sIdxThree,\sIdxFour}
			}
			{
				\ulDuration
				\normRhoP{\vIdxOne}{\vIdxTwo}^2
				\normRhoD{\vIdxOne}{\vIdxTwo}^2
				\normChannelPower_{\vIdxOne,\vIdxOne,\vIdxTwo}^2
			}
		}\!\!
	}^{-1}
	\label{eqn:sinrNonIterativeFiniteMEqvt}
	\end{align}
	\setcounter{equation}{28}
	\begin{flalign}
	\sinrUlNonIterative_{\vIdxOne,\vIdxTwo}
	&\geq
	\lrc
	{
		\sum\limits_{\sIdxOne=0}^{L-1}
		\sum\limits_{\sIdxTwo=0}^{K-1}
		\frac{\normRhoD{\sIdxOne}{\sIdxTwo}^2}{\ulDuration\normRhoP{\vIdxOne}{\vIdxTwo}^2\normRhoD{\vIdxOne}{\vIdxTwo}^2}
		+
		\frac{LK-1}{M\normRhoD{\vIdxOne}{\vIdxTwo}^2}
		+	
		\mathop
		{
			\sum\limits_{\sIdxOne=0}^{L-1}
			\sum\limits_{\sIdxTwo=0}^{K-1}
		}\limits_{\lrf{\sIdxOne\neq\vIdxOne,\sIdxTwo\neq\vIdxTwo}}			
		\mathop
		{
			\sum\limits_{\sIdxThree=0}^{L-1}
			\sum\limits_{\sIdxFour=0}^{K-1}
		}\limits_{\lrf{\sIdxThree\neq\sIdxOne,\sIdxFour\neq\sIdxTwo}}
		\frac{\normRhoD{\sIdxThree}{\sIdxFour}^2}
		{M\ulDuration\normRhoP{\vIdxOne}{\vIdxTwo}^2\normRhoD{\vIdxOne}{\vIdxTwo}^2}			
	}^{-1}&
	\label{eqn:sinrNonIterativeLowerBound}
	\end{flalign}
	\setcounter{equation}{\value{MYtempeqncnt}}
	\hrulefill
	\vspace*{-10pt}
\end{figure*}

\subsection{Power Control and Choice of Parameters $ \rhoP{\vIdxOne}{\vIdxTwo} $ and $ \rhoD{\vIdxOne}{\vIdxTwo} $}
From \eqref{eqn:sinrSuperimposedPilotNonIterativeInfM}, it can be seen that the SINR of a user is dependent on the product of the transmit powers and large-scale fading coefficients of the remaining $ LK-1 $ users in addition to the product of its own transmit power and large-scale fading coefficient. This dependence results in a situation similar to the near-far problem in code division multiple access (CDMA) systems, wherein users that have larger values of large-scale fading coefficient $ \beta $ swamp users that have smaller values of $ \beta $. Therefore, it becomes necessary to use power control to provide a uniform user experience. 

While the parameters $ \ulTotPower{\sIdxOne}{\sIdxTwo} $, $ \rhoD{\sIdxOne}{\sIdxTwo} $, and $ \rhoP{\sIdxOne}{\sIdxTwo} $ can be optimized by maximizing the sum-rate of all the users, i.e.,
\begin{align}
\mathop{\text{max}}_{\ulTotPower{\sIdxOne}{\sIdxTwo},\rhoD{\sIdxOne}{\sIdxTwo},\rhoP{\sIdxOne}{\sIdxTwo}} 
\lrf{
	\sum\limits_{\sIdxOne=0}^{L-1}
	\sum\limits_{\sIdxTwo=0}^{K-1} 
	\userUlRateSpNonIterative_{\sIdxOne,\sIdxTwo}
}
\label{eqn:powerControlIdealOptimizationProblem}
\end{align}
the optimization problem is in general non-convex and requires coordination between the BSs. As an alternative, a suboptimal solution that does not involve coordination between the BSs is obtained here for the parameters $ \ulTotPower{\sIdxOne}{\sIdxTwo} $, $ \rhoD{\sIdxOne}{\sIdxTwo} $, and $ \rhoP{\sIdxOne}{\sIdxTwo} $. This suboptimal solution will be shown to maximize a lower bound on the sum-rate, and it is as follows. 

The received signal in \eqref{eqn:superimposedPilotsDefn} can be equivalently written as
\begin{align}
\mbf{Y}_{\vIdxOne} 
&= 
\sum\limits_{\sIdxOne=0}^{L-1}
\sum\limits_{\sIdxTwo=0}^{K-1} 
\sqrt{\ulTotPower{\sIdxOne}{\sIdxTwo}}
\mbf{h}_{\vIdxOne,\sIdxOne,\sIdxTwo}
\lrc{
	\frac{\rhoD{\sIdxOne}{\sIdxTwo}}{\sqrt{\ulTotPower{\sIdxOne}{\sIdxTwo}}}
	\mbf{x}_{\sIdxOne,\sIdxTwo}^T
	+
	\frac{\rhoP{\sIdxOne}{\sIdxTwo}}{\sqrt{\ulTotPower{\sIdxOne}{\sIdxTwo}}}
	\spPilot{\sIdxOne}{\sIdxTwo}^T
}
\!\!
+ 
\!\!
\mbf{W}_{\vIdxOne}
\nonumber
\\
&=
\sum\limits_{\sIdxOne=0}^{L-1}
\sum\limits_{\sIdxTwo=0}^{K-1} 
\normChannelVector_{\vIdxOne,\sIdxOne,\sIdxTwo}
\lrc{
	\normRhoD{\sIdxOne}{\sIdxTwo}
	\mbf{x}_{\sIdxOne,\sIdxTwo}
	+
	\normRhoP{\sIdxOne}{\sIdxTwo}
	\spPilot{\sIdxOne}{\sIdxTwo}
}^T 
+ 
\mbf{W}_{\vIdxOne}
\label{eqn:superimposedPilotEqvtDefn}
\end{align}
where
\begin{align}
\normChannelVector_{\vIdxOne,\sIdxOne,\sIdxTwo}
&\triangleq
\sqrt{\ulTotPower{\sIdxOne}{\sIdxTwo}}
\mbf{h}_{\vIdxOne,\sIdxOne,\sIdxTwo}
\sim
\mathcal{CN}
\lrc{\mbf{0},\normChannelPower_{\vIdxOne,\sIdxOne,\sIdxTwo}\eye_{M}}
\\
\normChannelPower_{\vIdxOne,\sIdxOne,\sIdxTwo}
&\triangleq
\beta_{\vIdxOne,\sIdxOne,\sIdxTwo}
\;.\;
\ulTotPower{\sIdxOne}{\sIdxTwo}
\label{eqn:normChannelPowerDefn}
\\
\normRhoD{\sIdxOne}{\sIdxTwo}
&\triangleq
\sqrt{\frac{\rhoD{\sIdxOne}{\sIdxTwo}^2}{\ulTotPower{\sIdxOne}{\sIdxTwo}}}
>
0
\label{eqn:normRhoDDefn}
\\
\normRhoP{\sIdxOne}{\sIdxTwo}
&\triangleq
\sqrt{\frac{\rhoP{\sIdxOne}{\sIdxTwo}^2}{\ulTotPower{\sIdxOne}{\sIdxTwo}}}
>
0
\label{eqn:normRhoPDefn}
\\	
\normRhoP{\sIdxOne}{\sIdxTwo}^2 
&+ 
\normRhoD{\sIdxOne}{\sIdxTwo}^2
=
1\;.
\label{eqn:normRhoDnormRhoPRelation}
\end{align}
From \eqref{eqn:superimposedPilotEqvtDefn}, it can be seen that a system having arbitrary values of $ \beta_{\vIdxOne,\sIdxOne,\sIdxTwo} $, $ \ulTotPower{\sIdxOne}{\sIdxTwo} $, $ \rhoD{\sIdxOne}{\sIdxTwo} $, and $ \rhoP{\sIdxOne}{\sIdxTwo} $, can be reduced into an equivalent system with parameters $ \normChannelPower_{\vIdxOne,\sIdxOne,\sIdxTwo} $, $\normRhoD{\sIdxOne}{\sIdxTwo}$, and $ \normRhoP{\sIdxOne}{\sIdxTwo}$, such that \mbox{$ 0\leq\normRhoD{\sIdxOne}{\sIdxTwo},\normRhoP{\sIdxOne}{\sIdxTwo}\leq 1 $}. Substituting \eqref{eqn:normChannelPowerDefn} -- \eqref{eqn:normRhoDnormRhoPRelation} into \eqref{eqn:sinrNonIterativeFiniteM}, an equivalent expression for the SINR, as shown in \eqref{eqn:sinrNonIterativeFiniteMEqvt} (shown at the top of the next page) can be obtained. \setcounter{equation}{\value{equation}+1}

To obtain the parameter $ \ulTotPower{\sIdxOne}{\sIdxTwo} $, we propose using the statistics-aware power-control approach detailed in \cite{bjornson2015massive}, wherein user $ \vIdxTwo $ in cell $ \vIdxOne $ transmits at a power $ \ulTotPower{\vIdxOne}{\vIdxTwo} = \powerBackoff/\beta_{\vIdxOne,\vIdxOne,\vIdxTwo} $ where $ \powerBackoff $ is a design parameter. The parameter $ \powerBackoff $ is chosen such that the transmitted power from a user satisfies a maximum power constraint, and users with severely low SINRs that would need a transmit power larger than this constraint would be denied service. This power control policy results in an identical received power of $ \powerBackoff $ at the $ \vIdxOne\rth $ BS for all the users in cell $ \vIdxOne $.
In addition, as mentioned in \cite{bjornson2015massive}, the ratio $0\leq\beta_{\vIdxOne,\sIdxOne,\sIdxTwo}/\beta_{\sIdxOne,\sIdxOne,\sIdxTwo}\leq 1 $ is the relative strength of the interference received at BS $ \vIdxOne $ from a user in cell $ \sIdxOne $. This ratio is at most $ 1 $, when the user is at the edge of the $ \vIdxOne\rth $ cell, and reduces to zero as its distance from BS $ \vIdxOne $ increases. Therefore, setting $ \ulTotPower{\sIdxOne}{\sIdxTwo} = \powerBackoff/\beta_{\sIdxOne,\sIdxOne,\sIdxTwo} $ and using the definitions of $ \normChannelPower_{\vIdxOne,\sIdxOne,\sIdxTwo} $, $ \normRhoD{\sIdxOne}{\sIdxTwo} $, and $ \normRhoP{\sIdxOne}{\sIdxTwo} $, and the inequality $0\leq\beta_{\vIdxOne,\sIdxOne,\sIdxTwo}/\beta_{\sIdxOne,\sIdxOne,\sIdxTwo}\leq 1 $, the following equations can be obtained
\begin{align}
\normChannelPower_{\vIdxOne,\vIdxOne,\vIdxTwo} 
&=
\beta_{\vIdxOne,\vIdxOne,\vIdxTwo}
\;.\;
\ulTotPower{\vIdxOne}{\vIdxTwo}
=
\powerBackoff
\label{eqn:normChannelPower}
\\	
\normChannelPower_{\vIdxOne,\sIdxOne,\sIdxTwo} 
&=
\beta_{\vIdxOne,\sIdxOne,\sIdxTwo}
\;.\;
\ulTotPower{\sIdxOne}{\sIdxTwo}
\leq	
\beta_{\sIdxOne,\sIdxOne,\sIdxTwo}
\ulTotPower{\sIdxOne}{\sIdxTwo}
=
\powerBackoff
\quad
\forall
\;
\sIdxOne
\neq
\vIdxOne
\\
\rhoD{\vIdxOne}{\vIdxTwo}^2
\beta_{\vIdxOne,\vIdxOne,\vIdxTwo}
&=
\normRhoD{\vIdxOne}{\vIdxTwo}^2
\powerBackoff
\\
\rhoP{\vIdxOne}{\vIdxTwo}^2
\beta_{\vIdxOne,\vIdxOne,\vIdxTwo}
&=
\normRhoP{\vIdxOne}{\vIdxTwo}^2
\powerBackoff
\\
\rhoD{\sIdxOne}{\sIdxTwo}^2
\beta_{\vIdxOne,\sIdxOne,\sIdxTwo}
&\leq
\rhoD{\sIdxOne}{\sIdxTwo}^2
\beta_{\sIdxOne,\sIdxOne,\sIdxTwo}	
=
\normRhoD{\sIdxOne}{\sIdxTwo}^2
\powerBackoff
\quad
\forall
\;
\sIdxOne
\neq
\vIdxOne
\\
\rhoP{\sIdxOne}{\sIdxTwo}^2
\beta_{\vIdxOne,\sIdxOne,\sIdxTwo}
&\leq
\rhoP{\sIdxOne}{\sIdxTwo}^2
\beta_{\sIdxOne,\sIdxOne,\sIdxTwo}	
=
\normRhoP{\sIdxOne}{\sIdxTwo}^2
\powerBackoff
\quad
\forall
\;
\sIdxOne
\neq
\vIdxOne\;.
\end{align}
Substituting the above equations into \eqref{eqn:sinrNonIterativeFiniteMEqvt}, a lower bound on the SINR, as shown in \eqref{eqn:sinrNonIterativeLowerBound} (shown at the top of the page), can be obtained.\setcounter{equation}{\value{equation}+1}

However, the maximization of the lower bound on the SINR and hence, a lower bound on the sum rate, is still a non-convex problem in the parameters $ \normRhoD{\sIdxOne}{\sIdxTwo} $ and $ \normRhoP{\sIdxOne}{\sIdxTwo} $ and requires coordination between the BSs. To circumvent this problem, we restrict the parameters $ \normRhoD{\sIdxOne}{\sIdxTwo} $ and $ \normRhoP{\sIdxOne}{\sIdxTwo} $ such that $ \normRhoD{\sIdxOne}{\sIdxTwo} = \normRhoD{}{},\forall\;\sIdxOne,\sIdxTwo $ and  $ \normRhoP{\sIdxOne}{\sIdxTwo} = \normRhoP{}{},\forall\;\sIdxOne,\sIdxTwo $. The choice of this restriction is motivated by the observation from \eqref{eqn:normChannelPower} that the statistics-aware power control scheme results in the same large-scale path loss coefficient for all the desired users in the cell, irrespective of their locations. As a result, from the BS's perspective, each of its users are identical, and therefore, there is no benefit in assigning different values of $ \normRhoD{\sIdxOne}{\sIdxTwo} $ to different users. More importantly, such a restriction renders the choice of $ \normRhoD{}{}_{\mathrm{opt}} $ to depend only on $ L $, $ K $, $ \ulDuration $, and $ M $ as will be shown next. Setting $ \normRhoD{\sIdxOne}{\sIdxTwo} = \normRhoD{}{},\forall\;\sIdxOne,\sIdxTwo $ and $ \normRhoP{\sIdxOne}{\sIdxTwo} = \normRhoP{}{},\forall\;\sIdxOne,\sIdxTwo $ in \eqref{eqn:sinrNonIterativeLowerBound}, we obtain
\begin{align}
\sinrUlNonIterative_{\vIdxOne,\vIdxTwo}
\!
&\geq
\!\!
\left(
\!
\frac{LK}{\ulDuration\lrc{1-\normRhoD{}{}^2}}
\!
+
\!
\frac{1}{M}
\!
\lrc{
	\!\!
	\frac{LK-1}{\normRhoD{}{}^2}
	\!
	+
	\!
	\frac{\lrc{LK-1}^2}{\ulDuration\lrc{1-\normRhoD{}{}^2}}
	\!
}\!\!\!
\right)^{\!\!-1}\!\!.
\label{eqn:spSinrUlNonIterativeLowerBound}
\end{align}
Differentiating the right hand side of \eqref{eqn:spSinrUlNonIterativeLowerBound} with respect to $ \normRhoD{}{}^2 $ and setting the resulting expression to zero, the value of $ \normRhoD{}{}^2 $ that maximizes the lower bound on $ \sinrUlNonIterative_{\vIdxOne,\vIdxTwo} $ and the UL sum rate can be obtained as
\begin{align}
\normRhoD{}{}^2_{\mathrm{opt}}
\!\!
&= 
\!\!
\lrc{
	\!
	1
	\!
	+
	\!
	\sqrt{
			\!\frac{\frac{LK}{\ulDuration}
			\!+\!
			\frac{\lrc{LK-1}^2}{M\ulDuration} }{\frac{LK-1}{M}} } 
	\!}^{-1}
\!\!\!\!\!\!\!
\approx
\!
\lrc{
	\!
	1
	\!
	+
	\!
	\sqrt{\frac{M+LK}{\ulDuration}} }^{-1}
\label{eqn:normRhoDOptimal}
\end{align}
and the optimal value of $ \normRhoP{}{}^2 $ can be obtained as
\begin{align}
\normRhoP{}{}^2_{\mathrm{opt}}  
=
1 - \normRhoD{}{}^2_{\mathrm{opt}} 
\approx
\lrc{1+\sqrt{\frac{\ulDuration}{M+LK}} }^{-1}
\label{eqn:normRhoPOptimal}
\end{align}
where the approximations in \eqref{eqn:normRhoDOptimal} and \eqref{eqn:normRhoPOptimal} have been made assuming $ LK\gg 1 $ in order to obtain simpler expressions. 

Based on the fact established in this subsection that systems using $ \rhoD{}{},\rhoP{}{}, \beta,\mbf{h} $, and $ \normRhoD{}{},\normRhoP{}{}, \normChannelPower,\normChannelVector $ are equivalent, we drop the overbar for ease of notation and adopt the former set of symbols in the rest of the paper. In addition, we set \mbox{$ \ulTotPower{\sIdxOne}{\sIdxTwo} = \normRhoD{\sIdxOne}{\sIdxTwo}^2 + \normRhoP{\sIdxOne}{\sIdxTwo}^2 = 1, \; \forall \sIdxOne,\sIdxTwo $}. 
\renewcommand{\normRhoD}[2]{\rhoD{#1}{#2}}
\renewcommand{\normRhoP}[2]{\rhoP{#1}{#2}}

\subsection{Impact of $ \ulDuration $ on the Performance of Superimposed Pilots}
Using \eqref{eqn:sinrSuperimposedPilotNonIterativeInfM} and a fixed set of parameters $ \pilotReuseRatio $, $ \tau $, and $ K $, the following theorem presents an important condition that guarantees the superiority of methods based on superimposed pilots over the LS estimator that is based on time-multiplexed pilots.
\begin{theorem}
	\label{thm:kappa}
	With fixed values of $ K $, $ \pilotReuseRatio $, and $ \tau $ and if $ {M\rightarrow\infty} $, there exists a UL duration $ \thresholdUlDuration_{\vIdxOne,\vIdxTwo} $ beyond which a channel estimator based on superimposed pilots outperforms the LS based channel estimator that utilizes time-multiplexed pilots, in terms of the SINR performance, in any channel scenario $ \lrf{\beta_{\vIdxOne,\sIdxOne,\vIdxTwo}\;\vrule\; 0\leq\vIdxOne,\sIdxOne\leq L-1,\; 0\leq\vIdxTwo\leq K-1} $.
	\begin{proof}
		If $ \thresholdUlDuration_{\vIdxOne,\vIdxTwo} $ is defined as the number of symbols in the uplink such that 
		\eqref{eqn:sinrConventionalPilot} and \eqref{eqn:sinrSuperimposedPilotNonIterativeInfM} are equal, i.e.,
		\begin{align}
		\frac{\beta_{\vIdxOne,\vIdxOne,\vIdxTwo}^2}
		{
			\frac{1}{\thresholdUlDuration_{\vIdxOne,\vIdxTwo}\rhoP{\vIdxOne}{\vIdxTwo}^2\rhoD{\vIdxOne}{\vIdxTwo}^2}
			\sum\limits_{\sIdxOne=0}^{L-1}
			\sum\limits_{\sIdxTwo=0}^{K-1}
			\rhoD{\sIdxOne}{\sIdxTwo}^2
			\beta_{\vIdxOne,\sIdxOne,\sIdxTwo}^2
		}
		=
		\frac{\beta^2_{\vIdxOne,\vIdxOne,\vIdxTwo}}{\sum\limits_{\substack{\sIdxOne\neq \vIdxOne\\\sIdxOne\in\mathcal{L}_\vIdxOne(\pilotReuseRatio)}} \beta^2_{\vIdxOne,\sIdxOne,\vIdxTwo} } 
		\label{eqn:equateSinrSuperimposedConventionalPilot}
		\end{align}
		then it is evident from \eqref{eqn:sinrSuperimposedPilotNonIterativeInfM} and \eqref{eqn:equateSinrSuperimposedConventionalPilot} that $ \ulDuration > \thresholdUlDuration_{\vIdxOne,\vIdxTwo} $ is a sufficient condition for a method that is based on superimposed pilots to outperform the LS method that employs time-multiplexed pilots. In addition, $ \thresholdUlDuration_{\vIdxOne,\vIdxTwo} $ is given as
		\begin{equation}
		\thresholdUlDuration_{\vIdxOne,\vIdxTwo} 
		\triangleq 
		\frac{
			\frac{1}{\rhoP{\vIdxOne}{\vIdxTwo}^2\rhoD{\vIdxOne}{\vIdxTwo}^2}
			\sum\limits_{\sIdxOne=0}^{L-1}
			\sum\limits_{\sIdxTwo=0}^{K-1}
			\rhoD{\sIdxOne}{\sIdxTwo}^2
			\beta_{\vIdxOne,\sIdxOne,\sIdxTwo}^2
		}
		{
			\sum\limits_{\substack{\sIdxOne\neq \vIdxOne\\\sIdxOne\in\mathcal{L}_\vIdxOne(\pilotReuseRatio)}} \beta^2_{\vIdxOne,\sIdxOne,\vIdxTwo}
		}\;.
		\end{equation}
		This completes the proof.
	\end{proof}
\end{theorem}
\begin{remark}
	An important consequence of the above theorem is that in scenarios with negligible pilot contamination, the LS method based on superimposed pilots requires a large value of $ \ulDuration $ to outperform the LS method based on time-multiplexed pilots. As an example, consider the case when \mbox{$ \pilotReuseRatio = 1 $}, \mbox{$ \beta_{\vIdxOne,\vIdxOne,\vIdxTwo} = 1, \; \forall \vIdxTwo$}, \mbox{$ \beta_{\vIdxOne,\sIdxOne,\vIdxTwo} = \beta, \;\forall \sIdxOne\neq\vIdxOne,\vIdxTwo$}, and \mbox{$ \rhoD{\vIdxOne}{\vIdxTwo}^2 = \rhoP{\vIdxOne}{\vIdxTwo}^2,\forall\;\vIdxOne,\vIdxTwo $}. For such a scenario, $ \thresholdUlDuration_{\vIdxOne,\vIdxTwo} $ is given as 
	\begin{equation}
	\thresholdUlDuration_{\vIdxOne,\vIdxTwo} = 2K\lrc{1+\frac{1}{(L-1)\beta^2}} \ .
	\end{equation}
	Then, if the LS estimator based on superimposed pilots is required to maintain superiority over the LS estimator employing time-multiplexed pilots, $ \ulDuration $ must scale inversely with $ \beta^2 $. This dependence on $ \ulDuration $ is evident from the expression for the channel estimation error, which is given as
	\begin{align}
		\Delta\mbf{h}_{\vIdxOne,\vIdxOne,\vIdxTwo} 
		&\triangleq 
		\mbf{h}_{\vIdxOne,\vIdxOne,\vIdxTwo} 
		- 
		\widehat{\mbf{h}}_{\vIdxOne,\vIdxOne,\vIdxTwo} 
		= -\frac{1}{\ulDuration\rhoP{\vIdxOne}{\vIdxTwo}} 
		\nonumber\\
		&
		\times
		\lrc{
			\sum\limits_{\sIdxOne=0}^{L-1} \sum\limits_{\sIdxTwo=0}^{K-1} 
			\rhoD{\sIdxOne}{\sIdxTwo}
			\mbf{h}_{\vIdxOne,\sIdxOne,\sIdxTwo} 
			\mbf{x}_{\sIdxOne,\sIdxTwo}^T 
			+
			\mbf{W}_{\vIdxOne}
		}
		\mbf{p}_{\vIdxOne,\vIdxTwo}^* \ .
		\label{eqn:superimposedPilotNonIterativeEstimationError}
		\end{align}
\end{remark}
\begin{remark}
	We build upon the discussion in \cite{Marzetta2010Noncooperative} on grouping users based on their coherence times. While such a grouping does not offer any performance benefits to users when employing the approach in \cite{Marzetta2010Noncooperative}, the use of superimposed pilots offers low-mobility users an increase in throughput, by minimizing the channel estimation error resulting from transmitting the data alongside the pilots. This improvement in performance is a direct consequence of Theorem \ref{thm:kappa}.
\end{remark}
\begin{remark}
	
	The type of pilot transmitted by a user can also be chosen based on the coherence time. While users with high-mobility or low pilot contamination would find it sufficient to use time-multiplexed pilots, users with low-mobility who suffer from significant pilot contamination due to their proximity to the cell-edge or due to shadowing would significantly benefit from employing superimposed pilots. 
\end{remark}
\begin{remark}
	Superimposed pilots require coordination between BSs when assigning pilot sequences and synchronizing transmissions. In practical cellular networks, the cells are fairly large and therefore, it can be assumed that the interference is restricted to the first tier of cells and the interference from the second and higher tiers of cells can be neglected. Therefore, it is reasonable to assume that practical deployments of superimposed pilots will require pilot assignment only over the first tier of cells, implying that coordination is limited to only this first tier. This overhead is not very different from that required by time-multiplexed pilots in the presence of pilot reuse. The coordination and synchronization requirements of superimposed pilot-based systems that allocate pilots over the first tier of cells are similar to that of time-multiplexed pilot-based systems that have a pilot reuse factor of $ r=3 $ \cite{bjornson2015massive}.
\end{remark}
From \eqref{eqn:superimposedPilotNonIterativeEstimationError}, it can be seen that the error in the channel estimate includes interference resulting from transmitting data alongside the pilots. Hence, the quality of the channel estimate can be improved by eliminating the interference from the transmitted data through iterative data-aided schemes, thereby increasing the robustness of the proposed method with respect to $ \ulDuration $.



\section{Iterative Data-Aided Channel Estimation}
\label{sec:superimposedPilotsIterative}
In the iterative approach to channel estimation developed in this section, the estimated channel and data vectors of both the desired and interfering users are used in feedback in order to eliminate the first term in \eqref{eqn:superimposedPilotNonIterativeEstimationError}. In addition, to minimize error propagation between the channel estimates of different users, the iteration is started from the user with the highest SINR and is progressed in the decreasing order of the SINRs of the users. It has to be noted that the objective of this section is to demonstrate that iterative methods for channel estimation with superimposed pilots provide a significantly better SINR performance than their non-iterative counterparts, and hence we restrict ourselves to a simple iterative algorithm. However, there is scope for developing improved iterative algorithms in the future.

\subsection{Algorithm}
\label{sec:superimposedPilotIterativeAlgorithm}
For the sake of clarity and without loss of generality, we replace the two indices $ \sIdxTwo,\sIdxOne $ with a single index $ \sIdxThree $ that lies in the range $ 0\leq\sIdxThree\leq N-1 $, where $ N \triangleq KL $. The index $ \sIdxThree $ is used to index the users in all the $ L $ cells. In addition, we drop the index $ \vIdxOne $ and implicitly assume that the channel estimation is performed at the $ \vIdxOne\rth $ BS.  Then, \eqref{eqn:superimposedPilotsDefn} can be rewritten as
\begin{equation}
\mbf{Y} = \sum\limits_{\sIdxThree=0}^{N-1}\mbf{h}_{\sIdxThree} \lrc{\rhoD{}{\sIdxThree}\mbf{x}_{\sIdxThree}+\rhoP{}{\sIdxThree}\mbf{p}_{\sIdxThree}}^T + \mbf{W} \ .
\end{equation}
Since for large $ M $, the SINRs of the users are proportional to the users' path-loss coefficients, the users are arranged in the decreasing order of their path-loss coefficients, i.e., \mbox{$ \beta_{0} > \beta_{1} > \ldots > \beta_{N-1} $}.%
\footnote{It is assumed that the BSs have access to the exact values of the path-loss coefficients $ \beta_{\sIdxThree} $ and that there is no false-ordering. This assumption is reasonable since for large $ M $, the path-loss coefficients can be computed at the BS with negligible error by averaging the power of the channel coefficients over the entire array.
}
Then, using an estimate of $ \rhoD{}{\sIdxThree} \mbf{h}_{\sIdxThree} \mbf{x}_{\sIdxThree}^T \mbf{p}_{\sIdxThree}^* $ for each user as a correction factor to minimize the interference from other users, the corresponding channel estimate of user $ \vIdxTwo $ can be written as
\renewcommand{\sIdxTwo}{n}
\renewcommand{\sIdxThree}{k}
\begin{align}
\widehat{\mbf{h}}_{\vIdxTwo}^{(i)} 
&= 
\frac{1}{\ulDuration\rhoP{}{\vIdxTwo}} 
\left[\mbf{Y}
- 
\sum\limits_{\substack{\sIdxThree=0\\\sIdxThree\in\mathcal{U}_{\vIdxTwo}^{(i)}}}^{\vIdxTwo-1} 
\rhoD{}{\sIdxThree}
\widehat{\mbf{h}}_{\sIdxThree}^{(i)}
\lrc{\widehat{\mbf{x}}_{\sIdxThree}^{(i)}}^T 
\right. \nonumber \\
&\quad\left.-
\sum\limits_{\substack{\sIdxThree=\vIdxTwo\\\sIdxThree\in\mathcal{U}_{\vIdxTwo}^{(i)}}}^{N-1} 
\rhoD{}{\sIdxThree}
\widehat{\mbf{h}}_{\sIdxThree}^{(i-1)} \lrc{\widehat{\mbf{x}}_{\sIdxThree}^{(i-1)}}^T
\right] 
\mbf{p}_{\vIdxTwo}^*  
\label{eqn:iterativeChannelEstimate}
\end{align}
where $ \widehat{\mbf{h}}_{\vIdxTwo}^{(0)} = \mbf{0},\;\forall\;\vIdxTwo $ and $ \mathcal{U}_{\vIdxTwo}^{(i)} $ is the set of users whose estimated data is used in feedback in the $ i\rth $ iteration to estimate the channel vector of user $ \vIdxTwo $. The approach to obtain $ \mathcal{U}_{\vIdxTwo}^{\lrc{i}} $ has been detailed in Appendix \ref{appdx:iterativeChannelEstimationThreshold}, and involves selecting users such that the interference power, described in the next subsection, does not increase with each iteration. The channel estimate in the above equation is a modified version of the LS estimator defined in \eqref{eqn:superimposedPilotNonIterativeChannel} with an added correction factor. Utilizing the resulting channel estimate in an MF and decision operation, similar to \eqref{eqn:superimposedPilotNonIterativeMF} and \eqref{eqn:superimposedPilotNonIterativeHardSlicing}, the estimate of the data is obtained as follows
\begin{align}
\lrc{\widetilde{\mbf{x}}_{\vIdxTwo}^{(i)}}^T 
&= 
\frac{1}{M\rhoD{}{\vIdxTwo}\beta_{\vIdxTwo}} 
\lrc{\widehat{\mbf{h}}_{\vIdxTwo}^{(i)}}^H 
\lrc{\mbf{Y}
	- 
	\rhoP{}{\vIdxTwo}
	\widehat{\mbf{h}}_{\vIdxTwo}^{(i)}
	\mbf{p}_{\vIdxTwo}^T}
\label{eqn:iterativeDataEstimate}\\
\widehat{\mbf{x}}_{\vIdxTwo}^{(i)} 
&= 
\eta\lrc{\widetilde{\mbf{x}}_{\vIdxTwo}^{(i)}} \label{eqn:iterativeHardSlicedData}
\end{align}
where $ \widehat{\mbf{x}}_{\vIdxTwo}^{(0)} = \mbf{0},  \ \forall \ \vIdxTwo = 0,\ldots,N-1 $.
\begin{remark}
	If the unitary matrix $ \mathbf{P} $, whose columns are the superimposed pilots, is chosen as $ {\mathbf{P} = \mathrm{blkdiag}\{\mathbf{P}_0,\ldots,\mathbf{P}_{L-1}\}} $, where the $ \ell $'th block $ {\mathbf{P}_{\ell} \in \mathbb{C}^{K\times K}} $ is comprised of the orthogonal pilot sequences used by the $ K $ users in cell $ \ell $, then the latency introduced when the non-iterative method is employed is the same as that for time-multiplexed pilots. However, when the iterative method is employed, the channel and the data vectors of the users are required and therefore, the uplink data in the entire slot will have to be aggregated before estimating the channel, which introduces a latency of $ \ulDuration $ symbols.
\end{remark}
\begin{remark}
	From \eqref{eqn:superimposedPilotNonIterativeChannel}, the non-iterative method for channel estimation requires $ M\ulDuration $ operations per user, whereas the MF and decision operations in \eqref{eqn:superimposedPilotNonIterativeMF} and \eqref{eqn:superimposedPilotNonIterativeHardSlicing} require $ M $ and $ \ulDuration $ operations per user, respectively.
	
	For the iterative method with $ \nu $ iterations, the channel estimator, matched filter, and decision operations have a combined complexity of $ O( \nu M\ulDuration) + O (\nu M)+O(\nu\ulDuration) $.
\end{remark}

\subsection{Interference Power at the BS}
Let $ \xTildeError{\vIdxTwo}^{(i)} \triangleq \mbf{x}_{\vIdxTwo} - \widetilde{\mbf{x}}_{\vIdxTwo}^{(i)} $ be the error in the estimate of the data symbols of user $ \vIdxTwo $ obtained from the MF in the $ i\rth $ iteration. Let $ \Delta\mbf{x}_{\vIdxTwo}^{(i)} \triangleq \mbf{x}_\vIdxTwo - \widehat{\mbf{x}}_{\vIdxTwo}^{(i)} $ be the corresponding error vector after the decision operation and let
$ \Delta\mbf{h}^{(i)}_\vIdxTwo \triangleq \mbf{h}_{\vIdxTwo} - \widehat{\mbf{h}}_{\vIdxTwo}^{(i)} $ be the associated error in the channel estimate. If $ \alpha_{\sIdxTwo}^{(i)} $ is the variance of the elements of $ \Delta\mbf{x}_{\sIdxTwo}^{(i)} $ and assuming that the elements of $ \xTildeError{\vIdxTwo}^{(i)} $ are i.i.d. circular complex-Gaussian random variables with zero mean and variance $ \spUlInterferencePower_{\vIdxTwo}^{(i)} $,
an approximate expression for the interference power $ \spUlInterferencePower_{\vIdxTwo}^{(i)} $ can be written as
\renewcommand{\vIdxOne}{j}
\renewcommand{\vIdxTwo}{m}
\renewcommand{\sIdxOne}{\ell}
\renewcommand{\sIdxTwo}{n}
\renewcommand{\sIdxThree}{k}
\begin{align}
\spUlInterferencePower_{\vIdxTwo}^{(i)} 
&
\approx 
\frac{1}{\beta_{\vIdxTwo}^2}\!\!
\lrc{\!\!
	\frac{1}{M\rhoD{}{\vIdxTwo}^2} 
	\sum\limits_{\substack{\sIdxThree=0\\\sIdxThree\neq \vIdxTwo}}^{N-1}
	\beta_{\sIdxThree}
	\beta_{\vIdxTwo} 
	+
	\frac{\sigma^2\beta_{\vIdxTwo}}{M\rhoD{}{\vIdxTwo}^2}
	+
	\frac{1}{M^2\rhoD{}{\vIdxTwo}^2} 
	\psi_{\vIdxTwo}^{(i)}
	\!\!}
\label{eqn:interferenceExpression}
\end{align}
\renewcommand{\vIdxOne}{j}%
\renewcommand{\vIdxTwo}{m}%
\renewcommand{\sIdxOne}{\ell}%
\renewcommand{\sIdxTwo}{k}%
\renewcommand{\sIdxThree}{n}%
where the expression for $ \psi_{\vIdxTwo}^{(i)} $ is given in \eqref{eqn:psiDefn} on the top of the next page and $ \psi_{\vIdxTwo}^{(0)} = 0,\;\forall \vIdxTwo $. The detailed derivation of $ \spUlInterferencePower_{\vIdxTwo}^{(i)} $ can be found in Appendix \ref{appdx:sinrCalc}.

\begin{figure*}[!t]
	\normalsize
	\setcounter{MYtempeqncnt}{\value{equation}}
	\setcounter{equation}{41}
	\renewcommand{\vIdxOne}{j}
	\renewcommand{\vIdxTwo}{m}
	\renewcommand{\sIdxOne}{\ell}
	\renewcommand{\sIdxTwo}{n}
	\renewcommand{\sIdxThree}{k}
	\begin{align}
	&\psi_{\vIdxTwo}^{(i)}\bigg|_{i\geq 1}= 
	\frac{M^2}{\ulDuration\rhoP{}{\vIdxTwo}^2}
	\left[\sum\limits_{\sIdxThree\in\mathcal{U}_{\vIdxTwo}^{(i)},\sIdxThree<\vIdxTwo}
	\rhoD{}{\sIdxThree}^2
	\lrf{	
		\beta_{\sIdxThree}^2\alpha_{\sIdxThree}^{(i)}
		+
		\frac{1}{M}
		\sum\limits_{\sIdxTwo=0}^{N-1}
		\beta_{\sIdxTwo}
		\beta_{\sIdxThree}
		\alpha_{\sIdxThree}^{(i)}
		+
		\frac{\lrc{1+\alpha_{\sIdxThree}^{(i)}}}{M^2}
		\psi_{\sIdxThree}^{(i)}
	}
	+
	\sum\limits_{\sIdxThree\notin\mathcal{U}_{\vIdxTwo}^{(i)}}
	\rhoD{}{\sIdxThree}^2
	\left\{			
	\beta_{\sIdxThree}^2
	+
	\frac{1}{M}		
	\sum\limits_{\sIdxTwo=0}^{N-1}
	\beta_{\sIdxTwo}
	\beta_{\sIdxThree}
	\right
	\}
	\right. \nonumber\\
	&\left.			
	+	
	\sum\limits_{\sIdxThree\in\mathcal{U}_{\vIdxTwo}^{(i)},\vIdxTwo\leq\sIdxThree\leq N}
	\rhoD{}{\sIdxThree}^2
	\lrf{		
		\beta_{\sIdxThree}^2
		\alpha_{\sIdxThree}^{(i-1)}
		+
		\sum\limits_{\sIdxTwo=0}^{N-1}
		\frac{1}{M}
		\beta_{\sIdxTwo}
		\beta_{\sIdxThree}
		\alpha_{\sIdxThree}^{(i-1)}
		+
		\frac{\lrc{1+\alpha_{\sIdxThree}^{(i-1)}}}{M^2}
		\psi_{\sIdxThree}^{(i-1)}
	}			
	+
	\frac{\sigma^2}{M}
	\lrc{
		\sum\limits_{\sIdxTwo=0}^{N-1}
		\beta_{\sIdxTwo}	
	} \right] \;.
	\label{eqn:psiDefn}
	\end{align}
	\setcounter{equation}{\value{MYtempeqncnt}}
	\hrulefill
	\vspace*{-10pt}
\end{figure*}

\setcounter{equation}{\value{equation}+1}
In deriving \eqref{eqn:interferenceExpression}, the following simplifying assumptions have been made in order to obtain a closed form expression:
\begin{enumerate}[label=(S\arabic*),ref=(S\arabic*)]
	\item $ \xTildeError{\vIdxTwo}^{(i)} $ is independent of $ \mbf{x}_{\sIdxTwo}$ and $ \mbf{W}, \;\forall\;\sIdxTwo,i$.\label{S1}
	\item $ \Delta\mbf{x}_{\vIdxTwo}^{(i)} $ is independent of $ \mbf{x}_{\sIdxTwo}$, $ \mbf{W} $, and $ \mbf{h}_{\sIdxTwo}, \;\forall\; \sIdxTwo,i$. \label{S2}
	\item $ \Delta\mbf{x}_{\vIdxTwo}^{(i)} $ is independent of $ \Delta\mbf{x}_{\sIdxTwo}^{(p)},\;\forall p\neq i,\vIdxTwo\neq\sIdxTwo  $ and the elements of  $ \Delta\mbf{x}_{\vIdxTwo}^{(i)} $ are i.i.d. \label{S3}
	\item $ \Delta\mbf{h}_{\vIdxTwo}^{(i)} $ is independent of $ \mbf{x}_{\sIdxTwo} $, $ \mbf{W} $, and $ \Delta\mbf{x}_{\sIdxTwo}^{(p)}, \;\forall\; \sIdxTwo,p $.\label{S4}
\end{enumerate}
In scenarios with low interference and with large $ M $, only a few of the received symbols will be erroneous. As a result, the elements of $ \Delta\mbf{x}_{\vIdxTwo}^{(i)} $ are sparse  with the few non-zero elements restricted to locations that correspond to the erroneous symbols. Moreover, the vector $ \xTildeError{\vIdxTwo}^{(i)} $ represents the error in the estimated data and in such low-interference scenarios, the elements of $ \xTildeError{\vIdxTwo}^{(i)} $ take small values. Therefore, the simplifications \ref{S1}, \ref{S2}, and \ref{S3} are reasonably accurate for these scenarios. Although the expression for  $ \Delta\mbf{h}_{\vIdxTwo}^{(i)} $, (given in \eqref{eqn:deltaHdefn} in Appendix \ref{appdx:sinrCalc}) is explicitly dependent on $ \mbf{x}_{\sIdxTwo} $ and $ \Delta\mbf{x}_{\vIdxTwo}^{(i)} $, we neglect the correlation between these terms since $ \Delta\mbf{h}_{\vIdxTwo}^{(i)} $ is inversely proportional to $ \ulDuration $, and the simplification \ref{S4} is fairly accurate when $ \ulDuration $ is large with respect to $ N $ and when scenarios with low interference are considered. 
Since $ \xTildeError{\vIdxTwo}^{(i)} $ is assumed to be a zero-mean random variable, $ \Delta\mbf{x}_{\sIdxTwo}^{(i)} $ is also a zero-mean random variable, provided the constellation points in $ \chi $ and their probability density functions are symmetric about the origin. This is true since by definition, $ \Delta\mbf{x}_{\sIdxTwo}^{(i)} $ and $ \xTildeError{\vIdxTwo}^{(i)} $ are related to each other through the following equation
\begin{equation}
\Delta\mbf{x}_{\sIdxTwo}^{(i)} 
= 
\mbf{x}_{\sIdxTwo} 
- 
\eta\lrc{\mbf{x}_{\sIdxTwo}-\xTildeError{\vIdxTwo}^{(i)}} \;.
\label{eqn:superimposedPilotIterative_DeltaX_e_relation}
\end{equation}
From \eqref{eqn:superimposedPilotIterative_DeltaX_e_relation}, an expression for the variance of the elements of $ \Delta\mbf{x}_{\sIdxTwo}^{(i)} $, i.e., $ \alpha_{\sIdxTwo}^{(i)} $ can be found as
\begin{align}
\alpha_{\sIdxTwo}^{(i)} &\triangleq \mathbb{E}\lrf{\left|\lrs{\Delta\mbf{x}_{\sIdxTwo}^{(i)}}_n\right|^2} 
= 
\int
\left|\Delta x\right|^2 p_{\Delta\mbf{x}_{\sIdxTwo}^{(i)}}(\Delta x) d\Delta x  \nonumber\\
&=
\int\limits_{x\in\chi}\int
\left|
x
-
\eta\lrc{
	x 
	- 
	e
}\right|^2
p_{\xTildeError{\sIdxTwo}^{(i)},\mbf{x}_\sIdxTwo}
\lrc{
	e,x
}
de \; 
dx
\nonumber\\
&= 
\int\limits_{x\in\chi}\int
\left|
x
-
\eta\lrc{
	x 
	- 
	e
}\right|^2
p_{\xTildeError{\sIdxTwo}^{(i)}}
\lrc{
	e
}
p_{\mbf{x}_\sIdxTwo}
\lrc{x}
de\;
dx
\label{eqn:alphaDefn}
\end{align}
where $ p_{\xTildeError{\sIdxTwo}^{(i)}}(\cdot) $, $ p_{\Delta\mbf{x}_{\sIdxTwo}^{(i)}}(\cdot) $, and $ p_{\mbf{x}_\sIdxTwo}\lrc{\cdot} $ are the probability density functions of the elements of $ \xTildeError{\sIdxTwo}^{(i)} $, $ \Delta\mbf{x}_{\sIdxTwo}^{(i)} $, and $ \mbf{x}_\sIdxTwo $, respectively, and $ p_{\xTildeError{\sIdxTwo}^{(i)},\mbf{x}_\sIdxTwo} (\cdot)$ is the joint density function of the random variables $ \xTildeError{\sIdxTwo}^{(i)} $ and  $ \mbf{x}_\sIdxTwo $. The latter has been written as the product of their individual distributions in the final expression of \eqref{eqn:alphaDefn}, thanks to \ref{S1}.

\textit{Important example of $ \alpha_{\vIdxTwo}^{(i)} $: } When the elements of $ \mbf{x}_\vIdxTwo $ are uniformly distributed and take values from a unit-power $ P $-quarternary amplitude modulation (QAM) constellation, then under the assumption that the symbol errors in $ \Delta\mbf{x}_{\sIdxTwo}^{(i)} $ are dominated by the closest neighboring symbols, the expression for  $ \alpha_{\vIdxTwo}^{(i)} $ can be written as
\begin{align}
\alpha_{\vIdxTwo}^{(i)} = \begin{cases}
\frac{24}{\sqrt{P}\lrc{\sqrt{P}+1}}Q\lrc{\sqrt{\frac{\frac{3}{(P-1)}}{\spUlInterferencePower_{\vIdxTwo}^{(i)}}}}, & i\geq 1 \\
1, & i=0
\end{cases}
\label{eqn:alphaDefnPQAM}
\end{align}
where $ Q\lrc{\cdot} $ is the Q-function. The detailed derivation of the above expression can be found in Appendix \ref{appdx:alphaPQAMderivation}.

\section{Hybrid System}
\label{sec:hybridSystem}
\begin{figure}
	\centering
		\resizebox{0.5\textwidth}{!}{\includegraphics{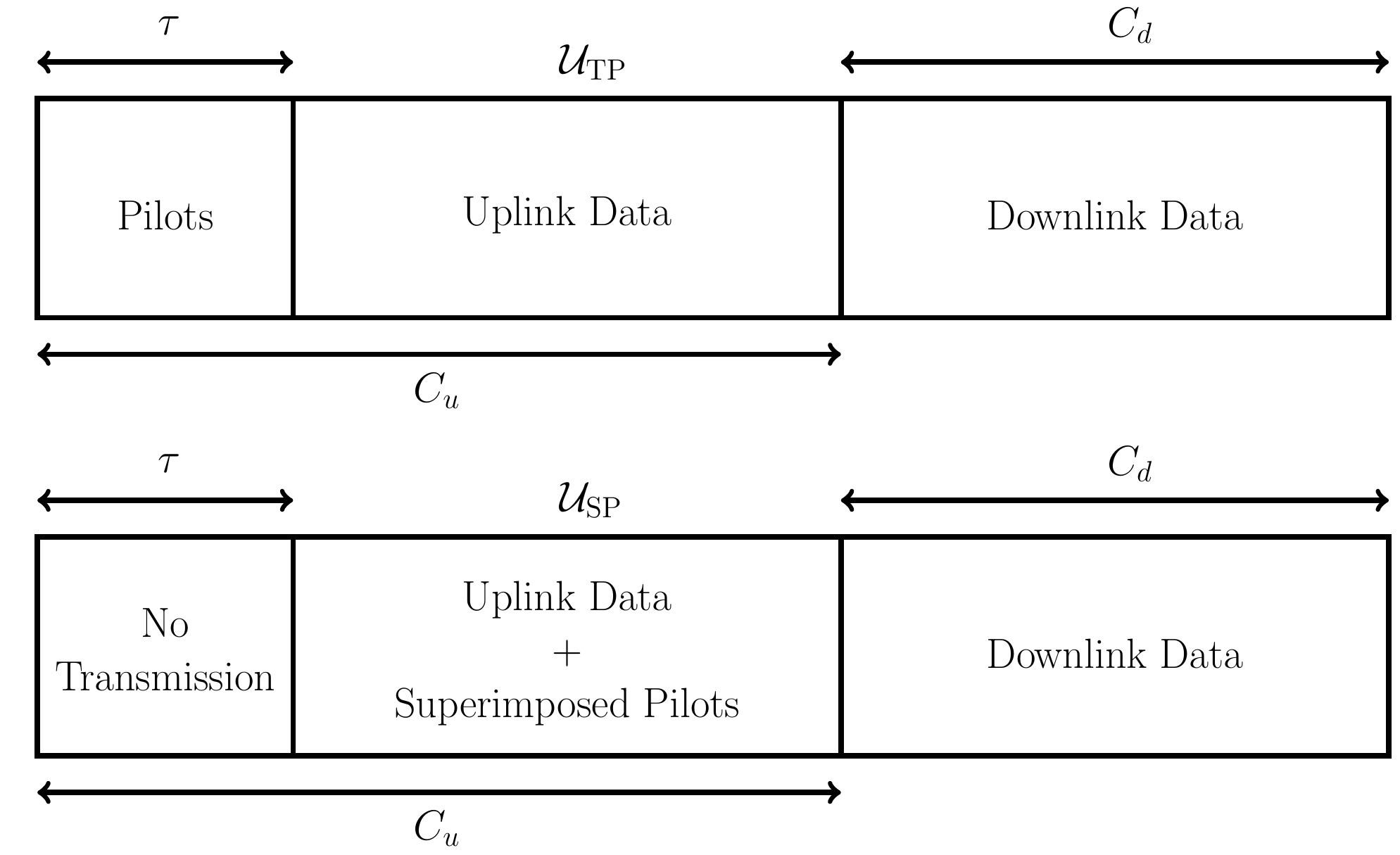}}	
	\vspace{-0.5cm}
	\caption{Frame structure of a hybrid system with users employing time-multiplexed and superimposed pilots.}
	\label{fig:hybridFrameStructure}
	\vspace{-0.25cm}
\end{figure}

One of the main advantages of superimposed pilots over time-multiplexed pilots is that it does not require a separate set of symbols for pilot transmission. This property can be used to construct a hybrid system that contains two disjoint sets of users, with the users in one of the sets employing time-multiplexed pilots, and the users in the other set employing superimposed pilots. The following theorem shows that this hybrid system has a higher throughput and supports a larger number of users than a system that employs only time-multiplexed pilots.

\begin{theorem}
	\label{thm:theoremHybridSystem}
	In a system that employs time-multiplexed pilots and is designed to maximize the UL and DL sum-rate,\footnote{Such as the scheme described in \textup{\cite{bjornson2015massive}.}} let $ K $ be the optimal number of users per cell, $ L $ be the total number of cells in the system, $ \tau>0 $ be the optimal number of symbols used for pilot training, $ \pilotReuseRatio $ be the optimal pilot-reuse factor, and $ \ulDuration -\tau  $ and $ \dlDuration $ be the number of data symbols in the UL and DL slots, respectively. Then, with $ M\rightarrow\infty $, there exists a hybrid system, that uses both time-multiplexed and superimposed pilots, which is capable of supporting $ \ulDuration -\tau $ additional users and offers a higher sum-rate in the UL than the optimal system that only employs time-multiplexed pilots.
	\begin{proof}
		Consider the frame structure in Fig. \ref{fig:hybridFrameStructure}, wherein there are two sets of users $ \cpUserSet $ and $ \spUserSet $. The users in the set $ \cpUserSet $ employ time-multiplexed pilots, with parameters selected using approaches such as in \cite{bjornson2015massive}. The users in the set $ \spUserSet $ maintain radio silence during the pilot training phase of the users in $ \cpUserSet $, i.e., for $ \tau $ symbols in the frame, and transmit orthogonal pilots superimposed with data during the uplink data phase of $ \ulDuration - \tau $ symbols. Since these users maintain radio silence during the pilot training phase of $ \tau $ symbols, they do not affect the quality of the channel estimates of the users in $ \cpUserSet $. As a result, under the assumption of asymptotic orthogonality of the channels, there is no interference from the users in $ \spUserSet $ to those in $ \cpUserSet $. Therefore, the per-cell sum-rate in the UL for the users in $ \cpUserSet $ remains unchanged and can be found from \eqref{eqn:rateConventionalPilots} to be
		\begin{align}
		\textrm{R}_{\vIdxOne}^{\mathrm{ul}}\lrc{\cpUserSet} = 
		\frac{(\ulDuration-\tau)}{\coherenceTime}
		\sum\limits_{\substack{\sIdxTwo=0\\\sIdxTwo\in\cpUserSet}}^{K-1}
		\log_2
		\lrc{
			1
			+
			\frac{\beta^2_{\vIdxOne,\vIdxOne,\sIdxTwo}}{\sum\limits_{\substack{\sIdxOne\neq \vIdxOne\\\sIdxOne\in\mathcal{L}_\vIdxOne(\pilotReuseRatio)}} \beta^2_{\vIdxOne,\sIdxOne,\sIdxTwo} }
		}\;.
		\label{eqn:rateUcp}
		\end{align}
		Assuming, for the sake of simplicity, that all the users in $ \spUserSet $ are located in the $ \vIdxOne\rth $ cell, the sum-rate of the users in $ \spUserSet $ can be found using \eqref{eqn:sinrSuperimposedPilotNonIterativeInfM} and \eqref{eqn:rateSuperimposedPilotNonIterative} as
		\begin{align}
		&\textrm{R}_{\vIdxOne}^{\mathrm{ul}}\lrc{\spUserSet} 
		= 
		\frac{(\ulDuration-\tau)}{\coherenceTime}
		\sum\limits_{\vIdxTwo\in\spUserSet}
		\log_2
		\lrc{
			1
			+
			\mathrm{SINR}_{\vIdxTwo}\lrc{\spUserSet}
		}
		\\
		&\mathrm{SINR}_{\vIdxTwo}\lrc{\spUserSet}
		\triangleq
		\frac{				
			\beta^2_{\vIdxOne,\vIdxOne,\vIdxTwo}
		}
		{
			\sum\limits_{\sIdxTwo\in\spUserSet}
			\frac{
				\rhoD{\vIdxOne}{\sIdxTwo}^2
				\beta_{\vIdxOne,\vIdxOne,\sIdxTwo}^2
			}
			{
				\lrc{\ulDuration-\tau}
				\rhoD{\vIdxOne}{\vIdxTwo}^2
				\rhoP{\vIdxOne}{\vIdxTwo}^2
			}
		}	\;.
		\label{eqn:rateUsp}
		\end{align}
		In obtaining the above expression, it has been assumed that the transmit power $ \ulTotPower{}{} $ of the users in $ \cpUserSet $ is small enough such that the interference to the users in $ \spUserSet $ can be neglected.\footnote{This assumption is valid since the SINR and the rate of the users in $ \cpUserSet $ are independent of the transmit power $ \ulTotPower{}{} $ when $ M\rightarrow \infty $. It has to be noted that this assumption has been made for the sake of simplicity and the theorem is valid even if this assumption does not hold.} Therefore, from \eqref{eqn:rateUcp} and \eqref{eqn:rateUsp}, the combined rate \mbox{$ \textrm{R}_{\vIdxOne}^{\mathrm{ul}}\lrc{\spUserSet} + \textrm{R}_{\vIdxOne}^{\mathrm{ul}}\lrc{\cpUserSet} $} is strictly greater than $ \textrm{R}_{\vIdxOne}^{\mathrm{ul}}\lrc{\cpUserSet} $. In addition, since the data slot is made up of $ \ulDuration-\tau $ symbols, it is possible to allocate \mbox{$ \ulDuration-\tau $} orthogonal pilots and therefore, the set $ \spUserSet $ can contain a maximum of $ \ulDuration-\tau $ users. This concludes the proof.
	\end{proof}
\end{theorem}
In the above theorem, given a system with users employing time-multiplexed pilots, we have shown that additional users employing superimposed pilots can always be added to the system, resulting in a hybrid system that offers a higher throughput. 

In the following section, we utilize the concept of the above theorem to partition a given set of users employing time-multiplexed pilots into two disjoint subsets $ \cpUserSet $ and $ \spUserSet $ that contain users transmitting time-multiplexed pilots and superimposed pilots, respectively. There are two main benefits of performing such a partition: (i) there is an overall improvement in the throughput as a result of the reduced inter-cell interference; and (ii) there is a reduction in the number of users that use time-multiplexed pilots, thereby allowing for more aggressive pilot reuse since $ \pilotReuseRatio $ is a function of the number of users employing time-multiplexed pilots \cite{bjornson2015massive}.


\section{A Simple Implementation of the Hybrid System}
\label{sec:hybridSystemImplementation}

Given a set of $ K $ users per cell in $ L $ cells with channel gains $ \beta_{\vIdxOne,\sIdxOne,\sIdxTwo}, \forall \vIdxOne,\sIdxOne = 1,\ldots,L,$ and $ \sIdxTwo=1,\ldots,K $, the problem of partitioning users into disjoint sets $ \cpUserSet $ and $ \spUserSet $ can be accomplished by minimizing the overall UL inter-cell and intra-cell interference. This choice of objective function is motivated by Theorem \ref{thm:kappa}, wherein it is observed that users at the cell edge cause significant pilot contamination and benefit from being assigned superimposed pilots, whereas users that are close to the BS cause negligible interference and could be assigned time-multiplexed pilots that are potentially shared with users in neighboring cells.

\subsection{Framework}
If the users in $ \cpUserSet $ transmit pilots with unit power and data at a power $ \ulTotPower{}{} $, then the received signal from the hybrid system in the UL phase at BS $ \vIdxOne $ can be written as
\begin{align}
\mbf{Y}_{\vIdxOne} = \mbf{Y}_{\vIdxOne}^{\mathrm{TP}} + \mbf{Y}_{\vIdxOne}^{\mathrm{SP}} + \mbf{W}_{\vIdxOne}
\end{align}
where $ \mbf{Y}_{\vIdxOne}^{\mathrm{TP}} $ and $ \mbf{Y}_{\vIdxOne}^{\mathrm{SP}} $ are the received signals from the users in $ \cpUserSet $ and $ \spUserSet $, respectively. From Fig. \ref{fig:hybridFrameStructure}, $ \mbf{Y}_{\vIdxOne}^{\mathrm{TP}} $ and $ \mbf{Y}_{\vIdxOne}^{\mathrm{SP}} $ can be written as
\begin{align}
\mbf{Y}_{\vIdxOne}^{\mathrm{TP}} 
&\triangleq 
\mathop{
	\sum\limits_{\sIdxOne=0}^{L-1}
	\sum\limits_{\sIdxTwo=0}^{K-1}
}_{\lrc{\sIdxOne,\sIdxTwo}\in\cpUserSet}
\mbf{h}_{\vIdxOne,\sIdxOne,\sIdxTwo}
\begin{bmatrix}	
\pmb{\phi}_{\sIdxOne,\sIdxTwo}^T , \sqrt{\ulTotPower{}{}}
\ulTxData_{\sIdxOne,\sIdxTwo}^T
\end{bmatrix}
\label{eqn:yCpDefn}
\\
\mbf{Y}_{\vIdxOne}^{\mathrm{SP}} 
&\triangleq 
\mathop{
	\sum\limits_{\sIdxOne=0}^{L-1}
	\sum\limits_{\sIdxTwo=0}^{K-1}
}_{\lrc{\sIdxOne,\sIdxTwo}\in\spUserSet}
\mbf{h}_{\vIdxOne,\sIdxOne,\sIdxTwo}
\begin{bmatrix}	
\mbf{0}_{1\times\tau} ,
\rhoD{}{}
\ulTxData_{\sIdxOne,\sIdxTwo}^T
+
\rhoP{}{}
\spPilot{\sIdxOne}{\sIdxTwo}^T
\end{bmatrix}
\end{align}
where the tuple $ \lrc{\sIdxOne,\sIdxTwo} $ is used to denote user $ \sIdxTwo $ in cell $ \sIdxOne $.

If user $ \lrc{\vIdxOne,\vIdxTwo}$ is a member of $ \cpUserSet  $, then the LS estimate of its channel can be written as \cite{Marzetta2010Noncooperative}
\begin{align}
\widehat{\mbf{h}}_{\vIdxOne,\vIdxOne,\vIdxTwo}
\!=\!
\frac{1}{\tau}
\mbf{Y}_{\vIdxOne}
\cpSelectionMatrix_{\vIdxOne,\vIdxTwo}
\!=\!
\mbf{h}_{\vIdxOne,\vIdxOne,\vIdxTwo}
\!
+
\!\!\!\!\!\!
\sum\limits_{\substack{\sIdxOne\neq\vIdxOne\\\sIdxOne\in\mathcal{L}_{\vIdxOne}(\pilotReuseRatio)\\\lrc{\sIdxOne,\vIdxTwo}\in\cpUserSet}}
\!\!\!\!\!\!
\mbf{h}_{\vIdxOne,\sIdxOne,\vIdxTwo}
+
\frac{1}{\tau}
\mbf{W}_{\vIdxOne}
\cpSelectionMatrix_{\vIdxOne,\vIdxTwo}
\label{eqn:hybridCpChannelEstimate}
\end{align}
where $ \cpSelectionMatrix_{\vIdxOne,\vIdxTwo} \triangleq \lrs{\pmb{\phi}^H_{\vIdxOne,\vIdxTwo} , \mbf{0}_{(1\times\lrc{\ulDuration-\tau})}}^T $.
If $ M\gg K $, the SINR in the UL when using the channel estimate in \eqref{eqn:hybridCpChannelEstimate} can be obtained similar to \eqref{eqn:sinrConventionalPilot} as
\begin{align}
\cpUlSinr_{\vIdxOne,\vIdxTwo}
&\approx
\frac
{
	\beta_{\vIdxOne,\vIdxOne,\vIdxTwo}^2
}
{
	\!\!\!\!\!\!\!
	\sum\limits_{\substack{\sIdxOne\neq\vIdxOne\\\sIdxOne\in\mathcal{L}_{\vIdxOne}(r)\\\lrc{\sIdxOne,\vIdxTwo}\in\cpUserSet}}
	\!\!\!\!\!\!\!
	\beta_{\vIdxOne,\sIdxOne,\vIdxTwo}^2
}
\label{eqn:hybridCpUlSinr}
\end{align}
where the approximations in \eqref{eqn:hybridCpUlSinr} is made for the sake of simplicity and is valid when $ M $ is sufficiently large. 

If user $ \lrc{\vIdxOne,\vIdxTwo} $ is a member of $ \spUserSet $, then the LS estimate of its channel can be written as
\begin{align}
&\widehat{\mbf{h}}_{\vIdxOne,\vIdxOne,\vIdxTwo}
=
\frac{1}{\lrc{\ulDuration-\tau}\rhoP{}{}}
\mbf{Y}_{\vIdxOne}
\spSelectionMatrix_{\vIdxOne,\vIdxTwo}
\nonumber
\\
&=
\mbf{h}_{\vIdxOne,\vIdxOne,\vIdxTwo}
+
\frac{\rhoD{}{}}{\lrc{\ulDuration-\tau}\rhoP{}{}}
\mathop{
	\sum\limits_{\substack{\sIdxOne=0}}^{L-1}
	\sum\limits_{\sIdxTwo=0}^{K-1}
}_{\lrc{\sIdxOne,\sIdxTwo}\in\spUserSet}
\mbf{h}_{\vIdxOne,\sIdxOne,\sIdxTwo}
\ulTxData_{\sIdxOne,\sIdxTwo}^T
\spPilot{\vIdxOne}{\vIdxTwo}^*
\nonumber
\\
&+	
\sqrt{\ulTotPower{}{}}
\mathop{
	\sum\limits_{\substack{\sIdxOne=0}}^{L-1}
	\sum\limits_{\sIdxTwo=0}^{K-1}
}_{\lrc{\sIdxOne,\sIdxTwo}\in\cpUserSet}
\frac
{
	\mbf{h}_{\vIdxOne,\sIdxOne,\sIdxTwo}
	\ulTxData_{\sIdxOne,\sIdxTwo}^T
	\spPilot{\vIdxOne}{\vIdxTwo}^*
}
{\lrc{\ulDuration-\tau}\rhoP{}{}}
+
\mbf{W}_{\vIdxOne}
\spSelectionMatrix_{\vIdxOne,\vIdxTwo}	
\label{eqn:hybridSpChannelEstimateInitial}
\end{align}
where $ \spSelectionMatrix_{\vIdxOne,\vIdxTwo} \triangleq \lrs{\mbf{0}_{(1\times \tau)} , \spPilot{\vIdxOne}{\vIdxTwo}^H}^T $.
Since it can be seen from \eqref{eqn:hybridCpUlSinr} that the UL SINR of the users in $ \cpUserSet $ is independent of the UL transmit power $ \ulTotPower{}{} $, we assume that $ \ulTotPower{}{} $ is small enough with respect to the transmit powers of the users in $ \spUserSet $.  As a result, the users in $ \spUserSet $ do not experience significant interference during the data transmission phase of the users in $ \cpUserSet $ and result in the transmissions of $ \spUserSet $ and $ \cpUserSet $ becoming independent of each other.\footnote{This assumption is made for the sake of clarity and simplicity. In the absence of this assumption, the BS will have to estimate and remove $ \mbf{Y}^{\mathrm{TP}}_{\vIdxOne} $ from $ \mbf{Y}_{\vIdxOne} $ before estimating the channels of the users in $ \spUserSet $.} Then \eqref{eqn:hybridSpChannelEstimateInitial} simplifies as
\begin{align}
\widehat{\mbf{h}}_{\vIdxOne,\vIdxOne,\vIdxTwo} 
&\approx
\mbf{h}_{\vIdxOne,\vIdxOne,\vIdxTwo}
+
\frac{\rhoD{}{}}{\rhoP{}{}}
\mathop{
	\sum\limits_{\substack{\sIdxOne=0}}^{L-1}
	\sum\limits_{\sIdxTwo=0}^{K-1}
}_{\lrc{\sIdxOne,\sIdxTwo}\in\spUserSet}
\frac{
	\mbf{h}_{\vIdxOne,\sIdxOne,\sIdxTwo}
	\ulTxData_{\sIdxOne,\sIdxTwo}^T
	\spPilot{\vIdxOne}{\vIdxTwo}^*
}
{\lrc{\ulDuration-\tau}}
+
\mbf{W}_{\vIdxOne}
\spSelectionMatrix_{\vIdxOne,\vIdxTwo}\;.
\label{eqn:hybridSpChannelEstimate}
\end{align}
Then the SINR in the UL for the users in $ \spUserSet $ can be obtained from \eqref{eqn:sinrSuperimposedPilotNonIterativeInfM} as
\begin{align}
\sinrUlNonIterative_{\vIdxOne,\vIdxTwo}
&\approx 
\frac{\beta_{\vIdxOne,\vIdxOne,\vIdxTwo}^2}
{
	\frac{1}{\lrc{\ulDuration-\tau}\rhoP{}{}^2}
	\underset{{\lrc{\sIdxOne,\sIdxTwo}\in\spUserSet}}
	{
		\sum\limits_{\sIdxOne=0}^{L-1}
		\sum\limits_{\sIdxTwo=0}^{K-1}
	}
	\beta_{\vIdxOne,\sIdxOne,\sIdxTwo}^2
}
\label{eqn:hybridSpUlSinr}
\end{align}
where, similar to \eqref{eqn:hybridCpUlSinr}, the approximation in \eqref{eqn:hybridSpUlSinr} is made for the sake of simplicity and is valid when $ M $ is sufficiently large.

\subsection{Algorithm to Obtain $ \cpUserSet $ and $ \spUserSet $.}
The goal in this subsection is to obtain an algorithm for partitioning users into the sets $ \cpUserSet $ and $ \spUserSet $ by minimizing the total UL inter-cell and intra-cell interference. In order to accomplish this, we quantify the amount of interference caused by a user that is assigned to either of the sets $ \cpUserSet $ or $ \spUserSet $. 

Let $ \hybridCpUlIci_{\vIdxOne,\vIdxTwo} $ or $ \hybridSpUlIci_{\vIdxOne,\vIdxTwo} $ be the contributions of user $ \lrc{\vIdxOne,\vIdxTwo} $ to the total UL inter/intra-cell interference power when assigned to $ \cpUserSet $ or $ \spUserSet $, respectively. If users $ \lrc{\vIdxOne,\vIdxTwo} $ and $ \lrc{\sIdxOne,\sIdxTwo} $ are members of $ \cpUserSet  $, then from the denominator of \eqref{eqn:hybridCpUlSinr}, the amount of interference that user $ \lrc{\vIdxOne,\vIdxTwo} $ causes to user $ \lrc{\sIdxOne,\sIdxTwo} $ in the UL is $ \beta_{\sIdxOne,\vIdxOne,\sIdxTwo}^2 \delta_{\vIdxTwo,\sIdxTwo} $.  Likewise, from \eqref{eqn:hybridSpUlSinr}, if both users are members of $ \spUserSet $, then the amount of interference that user $ \lrc{\vIdxOne,\vIdxTwo} $ causes to user $ \lrc{\sIdxOne,\sIdxTwo} $ in the UL is $ \beta_{\sIdxOne,\vIdxOne,\vIdxTwo}^2/\lrc{\lrc{\ulDuration-\tau}\rhoP{}{}^2} $. Therefore, $ \hybridCpUlIci_{\vIdxOne,\vIdxTwo} $ and $ \hybridSpUlIci_{\vIdxOne,\vIdxTwo} $ can be obtained as 
\begingroup
\allowdisplaybreaks
\begin{align}
\hybridCpUlIci_{\vIdxOne,\vIdxTwo} 
&=
\!\!\!\!
\underset{\substack{\sIdxOne\in\mathcal{L}_{\vIdxOne}(\pilotReuseRatio)\\\lrc{\sIdxOne,\sIdxTwo}\in\cpUserSet}}{
	\sum\limits_{\substack{\sIdxOne\neq\vIdxOne}}
	\sum\limits_{\sIdxTwo=0}^{K-1} 
}
\beta_{\sIdxOne,\vIdxOne,\sIdxTwo}^2
\delta_{\vIdxTwo,\sIdxTwo}
=
\!\!\!\!
\sum\limits_{\substack{\sIdxOne\neq\vIdxOne\\\sIdxOne\in\mathcal{L}_{\vIdxOne}(\pilotReuseRatio)\\\lrc{\sIdxOne,\vIdxTwo}\in\cpUserSet}}
\!\!\!\!\!\!\!
\beta_{\sIdxOne,\vIdxOne,\vIdxTwo}^2
\\
\hybridSpUlIci_{\vIdxOne,\vIdxTwo} 
&= 
\frac{1}{\lrc{\ulDuration-\tau}\rhoP{}{}^2}
\underset{{\lrc{\sIdxOne,\sIdxTwo}\in\spUserSet}}
{
	\sum\limits_{\sIdxOne=0}^{L-1}
	\sum\limits_{\sIdxTwo=0}^{K-1}
}
\beta_{\sIdxOne,\vIdxOne,\vIdxTwo}^2
\;.	
\end{align}
\endgroup
From the above equations, the total cost due to UL inter/intra-cell interference can be expressed as
\begin{align}
\hybridTotalIci\lrc{\cpUserSet,\spUserSet}
&= 
\sum\limits_{\sIdxOne=0}^{L-1}
\sum\limits_{\sIdxTwo=0}^{K-1}
\left(
\hybridCpUlIci_{\sIdxOne,\sIdxTwo}
\indicator{\lrc{\sIdxOne,\sIdxTwo}\in\cpUserSet}
\right.
\nonumber
\\
&\qquad\qquad
+
\left.
\hybridSpUlIci_{\sIdxOne,\sIdxTwo}
\indicator{\lrc{\sIdxOne,\sIdxTwo}\in\spUserSet}	
\right)
\label{eqn:hybridTotalIci}
\end{align}
Using \eqref{eqn:hybridTotalIci} as the objective function, the sets $ \cpUserSet $ and $ \spUserSet $ can be obtained as the solution of the following optimization problem
\begin{align}
\lrc{\cpUserSet,\spUserSet }
=
&\arg \min_{\substack{\cpUserSet\subseteq\hybridCompleteUserSet\\\spUserSet\subseteq\hybridCompleteUserSet}} \hybridTotalIci\lrc{\cpUserSet,\spUserSet} 
\nonumber
\\ 
&\text{subject to}\quad \cpUserSet\cup\spUserSet=\hybridCompleteUserSet 
\nonumber\\
&\qquad\qquad\quad\cpUserSet\cap\spUserSet=\varnothing
\label{eqn:hybridOptimizationProblem}
\end{align}
where $ \hybridCompleteUserSet $ is the set of all users in the $ L $ cells. However, the optimization problem in \eqref{eqn:hybridOptimizationProblem} is combinatorial in nature with $ 2^{\Card{\hybridCompleteUserSet}} $ possible choices for $ \cpUserSet $ and $ \spUserSet $, making it computationally hard to obtain the optimal solution. A workaround  is to employ a greedy approach to partition $ \hybridCompleteUserSet $ into $ \cpUserSet $ and $ \spUserSet $. At each step of this algorithm, given $ \cpUserSet $ and $ \spUserSet $, a user $ \lrc{\tilde{\sIdxOne},\tilde{\sIdxTwo}} $ in $ \cpUserSet $ is chosen as
\begin{equation}
\lrc{\tilde{\sIdxOne},\tilde{\sIdxTwo}} = \arg\max_{\lrc{\sIdxOne,\sIdxTwo}\in\cpUserSet} \hybridCpUlIci_{\sIdxOne,\sIdxTwo}\;.
\label{eqn:hybridGreedyAlgorithmWorstCaseUser}
\end{equation}
Setting $ \cpUserSetInterim = \cpUserSet\backslash\lrc{\tilde{\sIdxOne},\tilde{\sIdxTwo}} $ and $ \spUserSetInterim = \spUserSet\cup\lrc{\tilde{\sIdxOne},\tilde{\sIdxTwo}} $, user $ \lrc{\tilde{\sIdxOne},\tilde{\sIdxTwo}} $ is added to $ \spUserSet $ if
\begin{equation}
\hybridTotalIci\lrc{\cpUserSetInterim,\spUserSetInterim} \leq  \hybridTotalIci\lrc{\cpUserSet,\spUserSet}\;.
\label{eqn:hybridGreedyAlgorithmCompareStep}
\end{equation}
The algorithm is initialized with $ \cpUserSet = \hybridCompleteUserSet $ and is terminated when \eqref{eqn:hybridGreedyAlgorithmCompareStep} is no longer satisfied or when $ \cpUserSet $ is empty. The approach described above is summarized in Algorithm \ref{algo:hybridPartitioningAlgorithm}. 

The complexity of the greedy algorithm used for designing the hybrid system can be obtained as follows. The terms $ \hybridCpUlIci $ and $ \hybridSpUlIci $ require a maximum of $ \Card{\hybridCompleteUserSet} $ operations to compute, and therefore, computing $ \hybridTotalIci\lrc{\cpUserSet,\spUserSet} $ requires $ \Card{\hybridCompleteUserSet}^2 $  operations. Assuming that the greedy algorithm runs till the condition $ \cpUserSet = \varnothing $ is satisfied, then an upper bound on the computational complexity of the greedy algorithm is $ \Card{\hybridCompleteUserSet}^3 $ operations. Moreover, an overhead of $ 2\Card{\hybridCompleteUserSet} $ data transmissions is required for sending the large-scale path-loss coefficients to a central node and receiving the sets $ \cpUserSet $ and $ \spUserSet $.

It has to be noted that Algorithm \ref{algo:hybridPartitioningAlgorithm} is sub-optimal, but it is useful for illustrating the concept of the hybrid system. Partitioning algorithms that offer superior performance compared to Algorithm \ref{algo:hybridPartitioningAlgorithm} with lower coordination overhead are left as topics for future research.

\begin{algorithm}[t]	
	\begin{algorithmic}[1]
		\algnewcommand\algorithmicdata{\textbf{Data:}}
		\algnewcommand\data{\item[\algorithmicdata]}
		
		\algnewcommand\algorithmicinitialize{\textbf{Initialize:}}
		\algnewcommand\Initialize{\item[\algorithmicinitialize]}
		
		\data{$ \beta_{\vIdxOne,\sIdxOne,\sIdxTwo},\;\forall \vIdxOne,\sIdxOne=0,\ldots,L-1,\;\;\sIdxTwo=0,\ldots,K-1 $}
		\Initialize{$ \cpUserSet\leftarrow\hybridCompleteUserSet $, $ \spUserSet\leftarrow \varnothing $}
		\State Compute $ \lrc{\tilde{\sIdxOne},\tilde{\sIdxTwo}} $ as in \eqref{eqn:hybridGreedyAlgorithmWorstCaseUser} \label{algo:loopStart}
		\State Set $ \cpUserSetInterim \leftarrow \cpUserSet \backslash \lrc{\tilde{\sIdxOne},\tilde{\sIdxTwo}} $ and $ \spUserSetInterim \leftarrow \spUserSet \cup \lrc{\tilde{\sIdxOne},\tilde{\sIdxTwo}} $
		\If {$ \cpUserSet \neq \varnothing $ \textbf{and if} $ \hybridTotalIci\lrc{\cpUserSetInterim,\spUserSetInterim} \leq \hybridTotalIci\lrc{\cpUserSet,\spUserSet} $ }
		\State $ \cpUserSet := \cpUserSetInterim $, $ \spUserSet:= \spUserSetInterim $
		\State Return to Step \eqref{algo:loopStart}.
		\Else
		\State STOP			
		\EndIf
	\end{algorithmic}
	\caption{Greedy algorithm to select $ \cpUserSet $ and $ \spUserSet $ }\label{euclid}
	\label{algo:hybridPartitioningAlgorithm}		
\end{algorithm}




\newcommand{\plotScaleValue}{0.8}
\begin{figure*}[t!]
	\parbox{\textwidth}{
		\begin{minipage}[b]{0.48\textwidth}
			\centering
			\scalebox{0.8}{\includegraphics{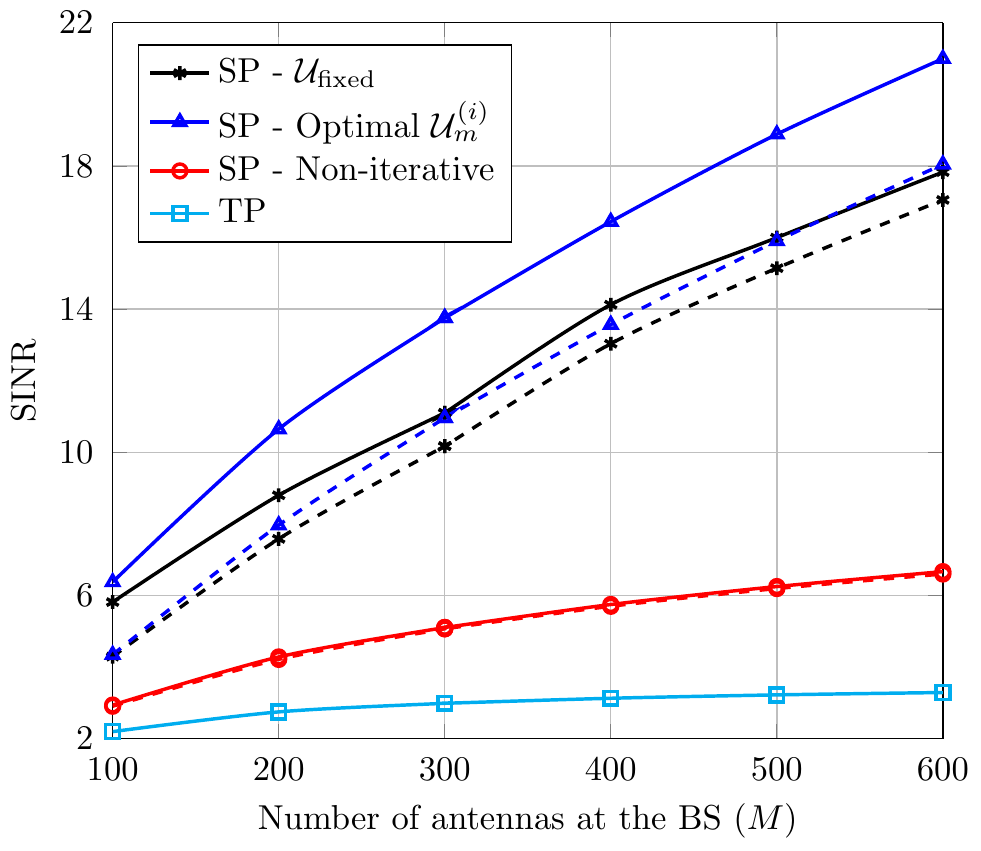}}
			\caption{The UL SINR of a user in the reference BS vs. $ M $ in Scenario~$ 2 $. The values of $ \normRhoD{}{} $ and $ \normRhoP{}{} $ are computed from \eqref{eqn:normRhoDOptimal} and \eqref{eqn:normRhoPOptimal}, respectively, and since they are approximations, they result in a non-smooth SINR behavior for the iterative methods. The solid and dashed lines represent simulated and theoretical curves, respectively.}
			\label{fig:SINRvsM-circular}
		\end{minipage}%
		\hspace{\stretch{2}}%
		\begin{minipage}[b]{0.48\textwidth}
			\centering
			\scalebox{0.8}{\includegraphics{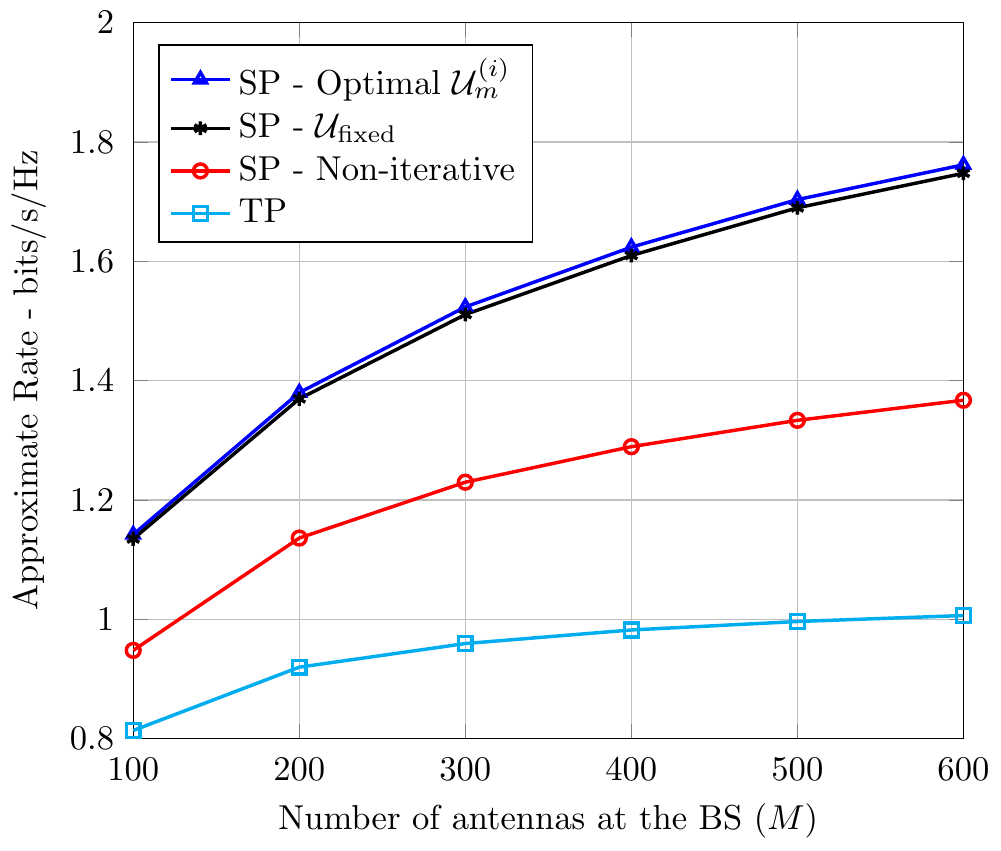}}
			\caption{Approximate per-user UL rate obtained using a $ 16 $-QAM constellation vs. $ M $ in the reference BS in Scenario~$ 2 $. The maximum UL rate that can be achieved with the $ 16$-QAM constellation, with half the symbols in a coherence blocked used for UL transmission, is $ 2 $ bps/Hz.}
			\label{fig:RatevsM-circular}
		\end{minipage}	
	}
	\vspace{-0.35cm}
\end{figure*}
\begin{figure*}[t!]
	\begin{minipage}[b]{0.48\textwidth}
		\centering
		\scalebox{\plotScaleValue}{\includegraphics{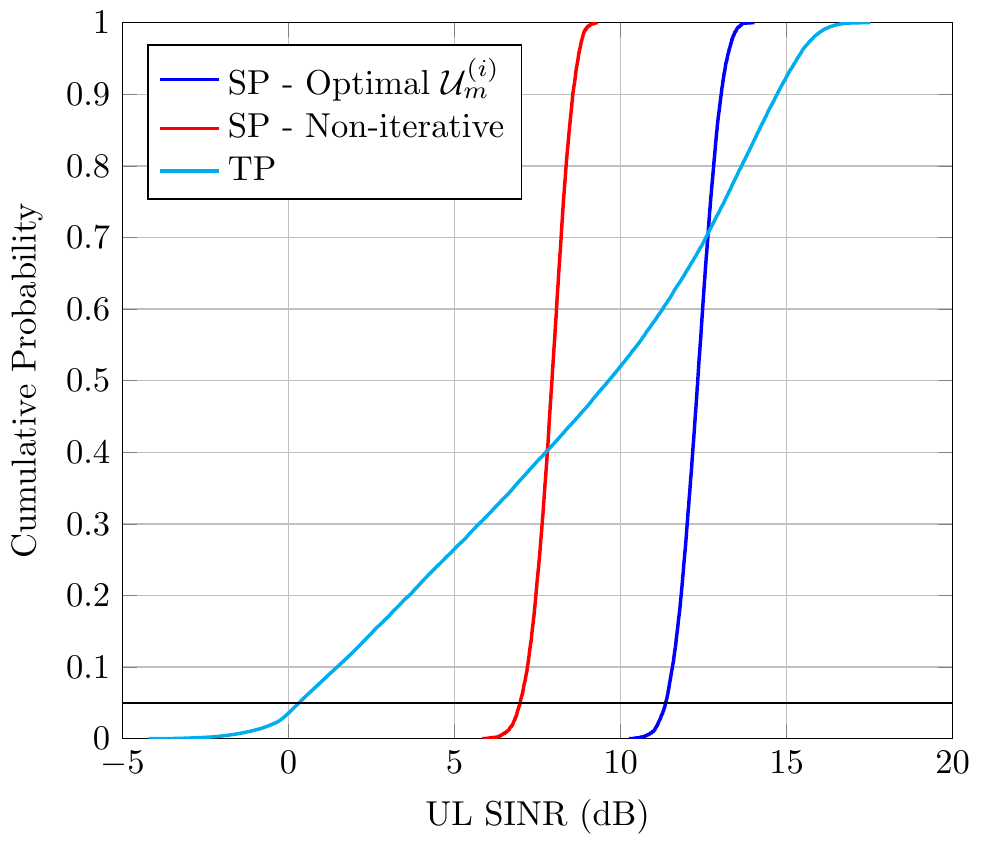}}		
		\caption{Cumulative distribution of the UL SINR in dBs for users in Scenario $ 1 $ with $ M=300 $ antennas. The black line indicates SINRs with probability $ \geq0.95 $.}
		\label{fig:SINR-CDF-geometric}
	\end{minipage}%
	\hspace{\stretch{2}}%
	\begin{minipage}[b]{0.48\textwidth}
		\centering
		\scalebox{\plotScaleValue}{\includegraphics{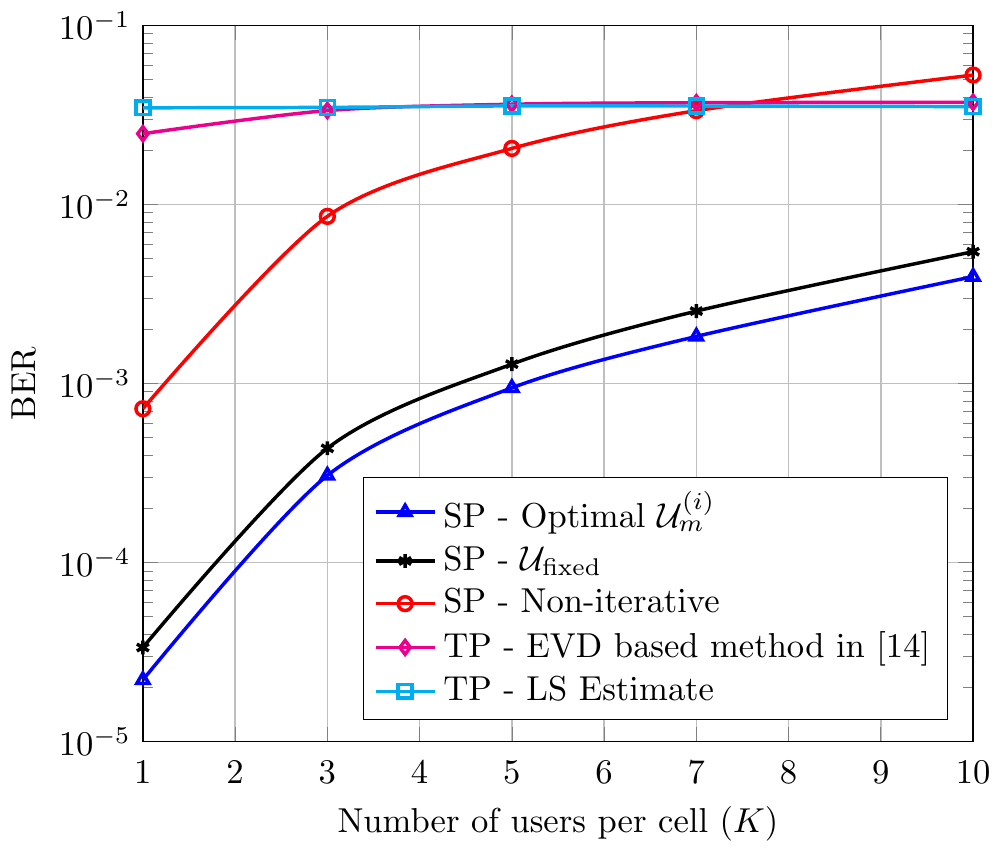}}		
		\caption{BER in the UL vs. $ K $ in Scenario~$ 1 $ with $ M/K=50 $ and $ \ulDuration = 70 $ symbols.}
		\label{fig:ulBerVsK}
	\end{minipage}%
	\vspace{-0.5cm}			
\end{figure*}

\section{Simulation Results}
\label{sec:numericalSimulations}

We compare the UL SINR and UL bit-error rate (BER) performance of the LS-based and eigenvalue decomposition (EVD)-based methods (referred to as `TP - EVD-based method' in the plots) that use time-multiplexed pilots to the performance of the channel estimator that uses superimposed pilots, at the output of a MF that employs these channel estimates. Two scenarios are considered for this comparison.
\begin{enumerate}
	\item[] \textit{Scenario $ 1 $}: The users are uniformly distributed in hexagonal cells of radius $ 1 $km with the BS at the center. In addition, users are located at a distance of at least $ 100 $m from the BS. 
	\item[] \textit{Scenario $ 2 $}: Users in both the reference and interfering cells are in a fixed configuration and are equally spaced on a circle of a given radius with the BS in the center. The size of the hexagonal cell is $ 1 $km and unless otherwise specified, the users are on a circle of radius $ 800 $m.
\end{enumerate}
Unless otherwise specified, the following parameters are used in both scenarios. The channel estimation methods are tested with $ L = 7 $ cells and $ K=5 $ users per cell. A $ P $-QAM constellation is employed and the path-loss coefficient is assumed to be $ 3 $. The simulations for the superimposed pilots-based iterative channel estimation scheme have been performed for $ 4 $ iterations. The number of symbols in the uplink time slot $ \ulDuration $ is set to $ 100 $, and for computing the rate, $ \coherenceTime $ is set to $ 200 $ symbols. The values of $ \normRhoD{}{} $ and $ \normRhoP{}{} $ are computed from \eqref{eqn:normRhoDOptimal} and \eqref{eqn:normRhoPOptimal}, respectively, and $ \powerBackoff $ is set to $ 1 $, where $ \powerBackoff $ is the design parameter in the statistics-aware power control scheme. The signal-to-noise ratio (SNR), i.e., $ \powerBackoff/\sigma^2 $ is set to $ 10 $dB. The methods based on time-multiplexed pilots have been simulated with $ \pilotReuseRatio = 1 $ and $ \ulTotPower{}{} = 1 $. In addition, the chosen channel estimation methods have been observed to perform better with the statistics aware-power control scheme, and therefore, this power control scheme has been employed for both time-multiplexed and superimposed pilots. The plots in Scenario~$ 1 $ are generated by averaging over $ 10^4 $ realizations of user locations across the cell. For each realization of user location, the channel vectors are generated and $ 200 $ bits are transmitted per user. The BER is computed by counting the bit errors for all the users in the reference cell. Similarly, the plots in Scenario~$ 2 $ are
generated for a fixed user location by averaging over $ 10^4 $ channel realizations with $ 200 $ bits transmitted per user for each realization.

Fig.~\ref{fig:SINRvsM-circular} shows the variation of the UL SINR of an arbitrary user with respect to $ M $ in Scenario~$ 2 $, whereas in Fig. \ref{fig:RatevsM-circular}, the approximate rate of an arbitrary user, calculated using $ 16$-QAM constellation, is plotted for the same scenario. We compute the achievable rate for 16-QAM signaling, modeling a practical scenario where highly mobile users are requesting moderate-to-high data rates. The SINR when the proposed method is employed, is shown to linearly increase in the number of antennas, whereas the SINR performance is observed to saturate for the LS-based method that uses time-multiplexed pilots. This trajectory of the proposed method could be potentially maintained using techniques such as adaptive modulation and coding, thereby implying that the effects of pilot contamination can be eliminated.

In Fig.~\ref{fig:SINR-CDF-geometric}, the cumulative distribution of the UL SINR in Scenario $ 1 $ is plotted. The interference power is averaged over $ 100 $ channel and data realizations for each realization of user location. While the LS-based method employing time-multiplexed pilots offers a higher SINR than the LS method employing superimposed pilots with a probability of approximately $ 0.6 $, the latter method can be seen to offer a significantly higher minimum SINR compared to the former method. Moreover, the users employing superimposed pilots have a smaller variation in their SINR than those employing time-multiplexed pilots. This is because the SINR of a user when superimposed pilots are employed is limited by the interference from the other users in the same cell, and the statistics-aware power control scheme renders the intra-cell interference power independent of the user location within the cell. The iterative method based on superimposed pilots is observed to offer a remarkably higher SINR performance with respect to its non-iterative counterpart and the LS-based method employing time-multiplexed pilots.

In Fig. \ref{fig:ulBerVsK}, the BER is plotted against the number of users per cell in Scenario~$ 1 $, with $ K $ ranging from $ 1 $ to $ 10 $ and $ \ulDuration = 70 $ symbols. Since $ L=7 $ cells, $ K=10 $ implies that the superimposed pilot-based system cannot support any new users without sharing pilots across cells. The ratio $ M/K $ is set to $ 50 $. While the non-iterative channel estimator based on superimposed pilots performs better in the UL at lower values of $ K $ than the estimators based on time-multiplexed pilots, the non-iterative estimator performs poorly at higher values of $ K $. This is because the data transmitted alongside the pilots causes self-interference and this interference power increases with the number of users in the system. Therefore, it is necessary to resort to iterative techniques to mitigate this additional interference and it can be seen that the iterative methods offer a better performance than methods based on time-multiplexed pilots when $ LK $ is close to $ \ulDuration $.

\begin{figure}[t!]
	\centering
	\scalebox{0.8}{\includegraphics{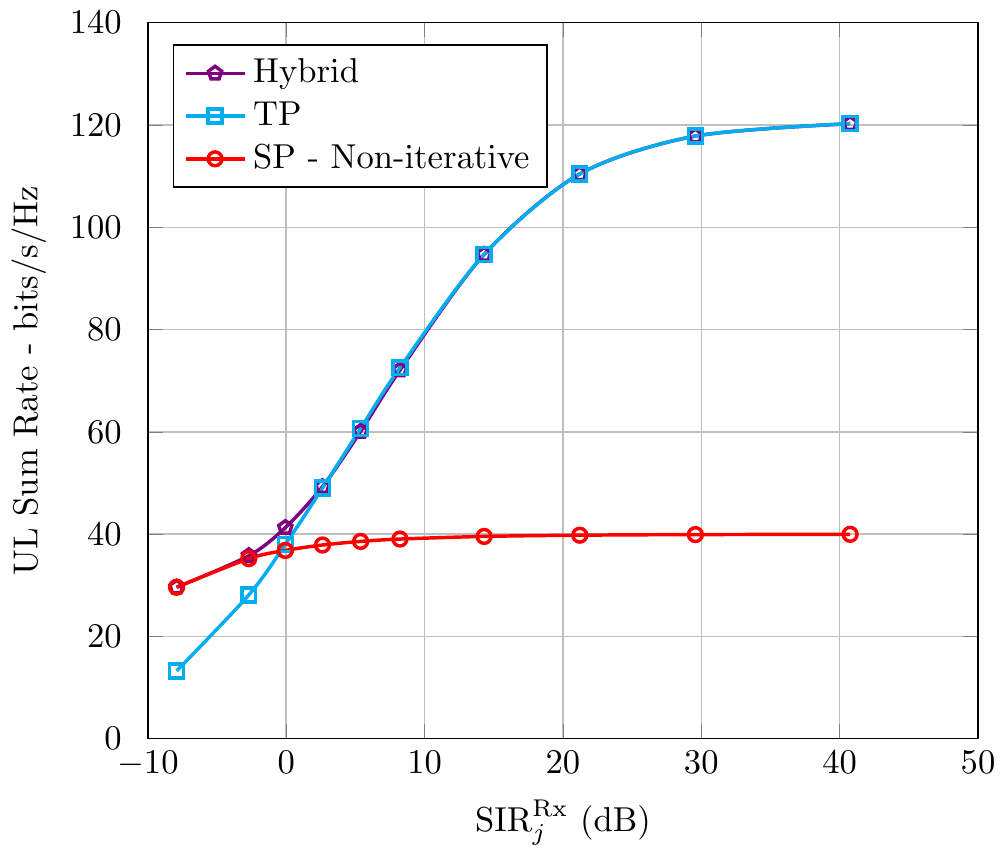}}	
	\caption{UL sum rate vs. $ \rxSinr_{\vIdxOne} $ in Scenario~$ 2 $ with $ M=1000 $ antennas.\newline}
	\label{fig:sumRateVsRadius-circular}		
	\vspace{-0.5cm}
\end{figure}

In Fig.~\ref{fig:sumRateVsRadius-circular}, the users are distributed as in Scenario~$ 2 $ and the distance of the users from the BS is varied between $ 0.2 $ and $ 0.9 $ km. For the chosen range of user distance, the total rate in the UL is plotted against the corresponding received signal-to-interference ratio (SIR). The received SIR of an arbitrary user $ \vIdxTwo $ in cell $ \vIdxOne $ is defined as
\begin{align}
\rxSinr_{\vIdxOne} 
\triangleq 
\frac{\powerBackoff }{ \sum\limits_{\sIdxOne\neq\vIdxOne}\sum\limits_{\sIdxTwo} \beta_{\vIdxOne,\sIdxOne,\sIdxTwo}^2}\;.
\end{align}
We assume $ L = 19 $ hexagonal cells, i.e., a central cell with two tiers of interfering cells. Each cell has $ M = 1000 $ antennas, $ K=5 $ users, and the value of $ \ulDuration $ is chosen as $ 40 $ symbols. Although $ L $ is set to $ 19 $, the optimization described in Algorithm \ref{algo:hybridPartitioningAlgorithm} and the computation of the performance metrics is performed over $ 7 $ cells which consist of the central and the first tier of cells. The value of $ \powerBackoff $ for users in $ \spUserSet $ is set to $ 10 $ and $ \ulTotPower{}{} $ for the users in $ \cpUserSet $ is set to $ 1 $. The data symbols are Gaussian distributed and the sum rate in Fig.~\ref{fig:sumRateVsRadius-circular} is obtained by averaging over $ 10^3 $ realizations of the channel and data symbols. 

In Fig.~\ref{fig:sumRateVsRadius-circular}, high and low values of SIR correspond to users located close to the BS and at the cell-edge, respectively. It can be observed that channel estimation methods based only on superimposed pilots (even the non-iterative formulation) are better in high interference scenarios, i.e., when the interfering users are at the cell-edge, whereas time-multiplexed pilots are better in low-interference scenarios. This behavior is a direct consequence of Theorem~\ref{thm:kappa} since higher interference scenarios have smaller values of $ \kappa $, resulting in superimposed pilots outperforming time-multiplexed pilots. However, at smaller values of user radius, the impact of pilot contamination is low but the self-interference in superimposed pilots resulting from transmitting the data alongside the pilots leads to a poorer performance compared to methods based on time-multiplexed pilots. In addition, it can be seen that the hybrid system adapts to the level of inter and intra-cell interference and offers a performance that is resilient to the location of the user within the cell. 

\section{Conclusion and Discussion}
\label{sec:conclusion}
We have proposed superimposed pilots as a superior alternative to time-multiplexed data and pilots for uplink channel estimation in massive MIMO. In the limit of an infinite number of antennas, a hybrid system using both superimposed pilots and time-multiplexed data and pilots offers a higher UL rate and supports larger number of users than the optimal system that utilizes only time-multiplexed data and pilots. The resilience to pilot contamination can be significantly improved with superimposed pilots through the use of an iterative data-aided channel estimation scheme that utilizes the data symbols of both the desired and interfering users in the feedback loop. Computer simulations in both a realistic scenario, in which users are distributed uniformly over the entire cell, and a high-interference scenario, in which users are concentrated at the cell edge, show that channel estimation methods using superimposed pilots offer a significant performance improvement over those that use time-multiplexed pilots.

The objective of this paper is to advocate superimposed pilots for practical use in massive MIMO systems by showing their superiority through theoretical and simulation based investigations. In standard MIMO communications, superimposed pilots are typically argued to be useful only for the scenario with high user mobility, and therefore, have not found practical application. On the contrary, in massive MIMO, superimposed pilots in a hybrid system provide superior performance in general. Therefore, there is a strong reason for superimposed pilots to make their way to practical use.

The proposed iterative data-aided channel estimation scheme and the greedy algorithm for partitioning users are suboptimal solutions to corresponding non-convex problems. Algorithms that offer performance close to the optimal solution at low computational complexities and overheads are of interest for future research. Moreover, the downlink performance of superimposed pilots is another topic of practical importance, which we have partially addressed in \cite{upadhya2016downlink}. 


\appendices
\section{}
\label{appdx:sinrNonIterative}
\renewcommand*{\sIdxOne}{\ell}
\renewcommand*{\sIdxTwo}{k}
\renewcommand*{\sIdxThree}{n}
\renewcommand*{\sIdxFour}{p}
\subsection*{Uplink SINR of the Non-Iterative Channel Estimation Method}
Using the notation described in Section \ref{sec:superimposedPilotIterativeAlgorithm}, \eqref{eqn:superimposedPilotsDefn} can be rewritten as
\begin{equation}
\mbf{Y} = \sum\limits_{\vIdxTwo=0}^{N-1}\mbf{h}_{\vIdxTwo} \lrc{\rhoD{}{\vIdxTwo}\mbf{x}_{\vIdxTwo}+\rhoP{}{\vIdxTwo}\mbf{p}_{\vIdxTwo}}^T + \mbf{W}
\end{equation}
From \eqref{eqn:superimposedPilotNonIterativeChannel}, the estimation error of the channel estimate can be obtained as
\begin{align}
\Delta\mbf{h}_{\vIdxTwo} 
&\triangleq 
\mbf{h}_{\vIdxTwo} 
- 
\widehat{\mbf{h}}_{\vIdxTwo} 
= -\frac{1}{\ulDuration\rhoP{}{\vIdxTwo}} \lrc{
	\sum\limits_{\sIdxTwo=0}^{N-1} 
	\rhoD{}{\sIdxTwo}
	\mbf{h}_{\sIdxTwo} 
	\mbf{x}_{\sIdxTwo}^T 
	+ 
	\mbf{W}
}
\mbf{p}_{\vIdxTwo}^* \ .
\label{eqn:superimposedPilotNonIterativeEstimationErrorAppdx}
\end{align}
From \eqref{eqn:superimposedPilotNonIterativeMF} and \eqref{eqn:superimposedPilotNonIterativeEstimationErrorAppdx}, the estimate of the received data after MF with the estimated channel can be written as
\begin{align}
&\widetilde{\mbf{x}}_{\vIdxTwo}^T
=
\frac{1}{M\rhoD{}{\vIdxTwo}\beta_{\vIdxTwo}}
\widehat{\mbf{h}}_{\vIdxTwo} ^H
\lrc{
	\mbf{Y}
	-
	\rhoP{}{\vIdxTwo}
	\widehat{\mbf{h}}_{\vIdxTwo} 
	\spPilot{}{\vIdxTwo}^T
}
\nonumber
\\
&= 
\frac{1}{M\rhoD{}{\vIdxTwo}\beta_{\vIdxTwo}} 
\lrc{\mbf{h}_{\vIdxTwo}^H - \Delta\mbf{h}_{\vIdxTwo}^H}\nonumber\\ &
\quad\times\left(
\sum\limits_{\sIdxTwo=0}^{N-1}
\mbf{h}_{\sIdxTwo}
\lrc{
	\rhoD{}{\sIdxTwo}
	\mbf{x}_{\sIdxTwo}
	+
	\rhoP{}{\sIdxTwo}
	\spPilot{}{\sIdxTwo}
}^T 
+ 
\mbf{W}
\right.\nonumber\\
&\quad\left.
-  
\rhoP{}{\vIdxTwo} 
\lrc{
	\mbf{h}_{\vIdxTwo} 
	- 
	\Delta\mbf{h}_{\vIdxTwo}
} 
\spPilot{}{\vIdxTwo}^T
\right)
=
\mbf{g}^T
+
\mbf{i}^T
\label{eqn:superimposedPilotNonIterativeSINRInterimStep}
\end{align}
where $ \mbf{g} $ and $ \mbf{i} $ are the signal and interference components of the matched filtered signal, respectively, which can be written as
\begingroup
\allowdisplaybreaks
\begin{align}
\mbf{g}
&\triangleq
\frac{\|\mbf{h}_{\vIdxTwo}\|^2}{M\beta_{\vIdxTwo}}				
\mbf{x}_{\vIdxTwo}
\\
\mbf{i}
&\triangleq
\sum\limits_{\sIdxThree=1}^{5}
\mbf{i}_{\sIdxThree}
\\
\mbf{i}_1
&\triangleq
\sum\limits_{\substack{\sIdxThree=0\\\sIdxThree\neq\vIdxTwo}}^{N-1}
\frac{\mbf{h}_{\vIdxTwo}^H\mbf{h}_{\sIdxThree}}{M\rhoD{}{\vIdxTwo}\beta_{\vIdxTwo}}
\lrc{
	\rhoP{}{\sIdxThree}
	\spPilot{}{\sIdxThree}
	+
	\rhoD{}{\sIdxThree}
	\mbf{x}_{\sIdxThree}
}
+
\frac
{\lrc{
		\mbf{h}_{\vIdxTwo}^H
		\mbf{W}
	}^T}
{M\rhoD{}{\vIdxTwo}\beta_{\vIdxTwo}}
\label{eqn:t1defn}
\\
\mbf{i}_2
&\triangleq
\frac{\rhoP{}{\vIdxTwo}}{M\rhoD{}{\vIdxTwo}\beta_{\vIdxTwo}}
\mbf{h}_{\vIdxTwo}^H		
\Delta\mbf{h}_{\vIdxTwo}
\spPilot{}{\vIdxTwo}
\\
\mbf{i}_3
&\triangleq
-
\frac{1}{M\beta_{\vIdxTwo}}
\Delta\mbf{h}_{\vIdxTwo}^H
\mbf{h}_{\vIdxTwo}
\mbf{x}_{\vIdxTwo}
\\
\mbf{i}_4
&\triangleq
-
\frac{1}{M\rhoD{}{\vIdxTwo}\beta_{\vIdxTwo}}		
\sum\limits_{\substack{\sIdxThree=0\\\sIdxThree\neq\vIdxTwo}}^{N-1}
\Delta\mbf{h}_{\vIdxTwo}^H
\mbf{h}_{\sIdxThree}
\lrc{
	\rhoP{}{\sIdxThree}
	\spPilot{}{\sIdxThree}
	+
	\rhoD{}{\sIdxThree}
	\mbf{x}_{\sIdxThree}
}
-
\frac{
	\lrc{
		\Delta\mbf{h}_{\vIdxTwo}^H
		\mbf{W}
	}^T	
}
{M\rhoD{}{\vIdxTwo}\beta_{\vIdxTwo}}
\\
\mbf{i}_5
&\triangleq
-
\frac{\rhoP{}{\vIdxTwo}}{M\rhoD{}{\vIdxTwo}\beta_{\vIdxTwo}}
\|\Delta\mbf{h}_{\vIdxTwo}\|^2
\spPilot{}{\vIdxTwo}
\;.
\label{eqn:t5defn}
\end{align}
\endgroup
The average interference power can be found as
\begin{align}
\expectation
\lrf{
	\|\mbf{i}\|^2
}
&=
\expectation
\lrf{
	\left\|
	\sum\limits_{\sIdxThree=1}^{5}
	\mbf{i}_{\sIdxThree}
	\right\|^2
}\;.
\label{eqn:interferencePowerEqn}
\end{align}
Then, using the definitions of $ \mbf{i}_{\sIdxThree},\forall\;\sIdxThree $ in \eqref{eqn:t1defn} -- $ \eqref{eqn:t5defn} $ and the definition of $ \Delta\mbf{h}_{\vIdxTwo} $ in \eqref{eqn:superimposedPilotNonIterativeEstimationErrorAppdx}, the following expressions can be easily obtained
\begingroup
\allowdisplaybreaks
\begin{align}
&\expectation
\lrf{\|\mbf{i}_1\|^2}
\approx
\frac{\ulDuration}{M\rhoD{}{\vIdxTwo}^2\beta_{\vIdxTwo}}
\sum\limits_{\substack{\sIdxThree=0\\\sIdxThree\neq\vIdxTwo}}^{N-1}
\beta_{\sIdxThree}
\ulTotPower{}{\sIdxThree}
\label{eqn:t1power}
\\
&\expectation
\lrf{\|\mbf{i}_2\|^2 + \|\mbf{i}_3\|^2 + \|\mbf{i}_4\|^2}
\approx
\sum\limits_{\substack{\sIdxThree=0\\\sIdxThree\neq\vIdxTwo}}^{N-1}
\sum\limits_{\substack{\sIdxTwo=0\\\sIdxTwo\neq\sIdxThree}}^{N-1}
\frac{\rhoD{}{\sIdxTwo}^2\beta_{\sIdxThree}\beta_{\sIdxTwo}\ulTotPower{}{\sIdxThree}}{M\rhoP{}{\vIdxTwo}^2\rhoD{}{\vIdxTwo}^2\beta_{\vIdxTwo}^2}
\!
\nonumber
\\
&
+
\sum\limits_{\sIdxThree=0}^{N-1}
\!\!
\frac{\rhoD{}{\sIdxThree}^2\ulTotPower{}{\sIdxThree}\beta_{\sIdxThree}^2}{\rhoP{}{\vIdxTwo}^2\rhoD{}{\vIdxTwo}^2\beta_{\vIdxTwo}^2}
+
\mathop
{
	\sum\limits_{\substack{\sIdxThree=0\\\sIdxThree\neq\vIdxTwo}}^{N-1}
	\sum\limits_{\substack{\sIdxFour=0\\\sIdxFour\neq\vIdxTwo}}^{N-1}
}_{\sIdxThree\neq\sIdxFour}
\frac{\rhoD{}{\sIdxThree}^2\rhoD{}{\sIdxFour}^2\beta_{\sIdxFour}\beta_{\sIdxThree}}{\ulDuration\rhoD{}{\vIdxTwo}^2\rhoP{}{\vIdxTwo}^2\beta_{\vIdxTwo}^2}
\\
&\expectation
\lrf{\|\mbf{i}_5\|^2}
\approx
\mathop{\sum\limits_{\substack{\sIdxThree=0\\\sIdxThree\neq\vIdxTwo}}^{N-1}
	\sum\limits_{\substack{\sIdxFour=0\\\sIdxFour\neq\vIdxTwo}}^{N-1}}_{\sIdxThree\neq\sIdxFour}
\frac{\rhoD{}{\sIdxThree}^2\rhoD{}{\sIdxFour}^2\beta_{\sIdxThree}\beta_{\sIdxFour}}{\ulDuration \rhoP{}{\vIdxTwo}^2\rhoD{}{\vIdxTwo}^2\beta_{\vIdxTwo}^2}
\\
&\expectation
\lrf{\lrc{\mbf{i}_3^H+\mbf{i}_4^H}\mbf{i}_5}
\approx
-
\mathop{
	\sum\limits_{\substack{\sIdxFour=0\\\sIdxFour\neq\vIdxTwo}}^{N-1}
	\sum\limits_{\substack{\sIdxThree=0\\\sIdxThree\neq\vIdxTwo}}^{N-1}
}_{\sIdxThree\neq\sIdxFour}
\frac{\rhoD{}{\sIdxThree}^2\rhoD{}{\sIdxFour}^2\beta_{\sIdxThree}\beta_{\sIdxFour}}{\ulDuration\rhoP{}{\vIdxTwo}^2\rhoD{}{\vIdxTwo}^2\beta_{\vIdxTwo}^2}
\label{eqn:t5t6t8power}
\end{align}
\endgroup
where the approximation errors in {\eqref{eqn:t1power}~--~\eqref{eqn:t5t6t8power}} are proportional to either $ N/M $, $ N/\ulDuration $, or $ \ulDuration/M $. In addition, the remaining terms of the form $ \mbf{i}_{\sIdxThree}^H\mbf{i}_{\sIdxFour},\forall\;\sIdxThree\neq\sIdxFour $ in the expansion of \eqref{eqn:interferencePowerEqn} are proportional to $ N/M $ or $ N/\ulDuration $. If $ M $ is large with respect to $ N $ and $ \ulDuration $, then the approximation errors and terms that are proportional to $ {N/M} $ and $ N/\ulDuration $ can be neglected. Similarly, error terms that are proportional to $ N/\ulDuration $ can also be dropped, and if $ \sigma^2 \ll \ulDuration $, then the effect of noise can also be neglected. Then, substituting {\eqref{eqn:t1power}~--~\eqref{eqn:t5t6t8power}} into the expansion of \eqref{eqn:interferencePowerEqn}, the interference power is obtained as
\begin{align}
\expectation
\lrf{
	\|\mbf{i}\|^2
}
&\approx
\sum\limits_{\sIdxThree=0}^{N-1}
\frac{\rhoD{}{\sIdxThree}^2\ulTotPower{}{\sIdxThree}\beta_{\sIdxThree}^2}{\rhoP{}{\vIdxTwo}^2\rhoD{}{\vIdxTwo}^2\beta_{\vIdxTwo}^2}
+
\sum\limits_{\substack{\sIdxThree=0\\\sIdxThree\neq\vIdxTwo}}^{N-1}
\frac{\ulDuration\beta_{\sIdxThree}\ulTotPower{}{\sIdxThree}}{M\rhoD{}{\vIdxTwo}^2\beta_{\vIdxTwo}}
\nonumber
\\
&
+
\sum\limits_{\substack{\sIdxThree=0\\\sIdxThree\neq\vIdxTwo}}^{N-1}
\sum\limits_{\substack{\sIdxTwo=0\\\sIdxTwo\neq\sIdxThree}}^{N-1}
\frac{\rhoD{}{\sIdxTwo}^2\beta_{\sIdxThree}\beta_{\sIdxTwo}\ulTotPower{}{\sIdxThree}}{M\rhoP{}{\vIdxTwo}^2\rhoD{}{\vIdxTwo}^2\beta_{\vIdxTwo}^2}
\label{eqn:InterferencePowerNonIterative}
\;.
\end{align}
Using \eqref{eqn:InterferencePowerNonIterative}, the SINR can be obtained as
\begin{align}
&\sinrUlNonIterative_{\vIdxTwo}
\triangleq
\frac{\expectation\lrf{\|\mbf{g}\|^2}}{\expectation\lrf{\|\mbf{i}\|^2}}
\nonumber
\\
&=
\frac{\ulDuration}
{
	\sum\limits_{\sIdxThree=0}^{N-1}
	\frac{\rhoD{}{\sIdxThree}^2\ulTotPower{}{\sIdxThree}\beta_{\sIdxThree}^2}{\rhoP{}{\vIdxTwo}^2\rhoD{}{\vIdxTwo}^2\beta_{\vIdxTwo}^2}
	+
	\sum\limits_{\substack{\sIdxThree=0\\\sIdxThree\neq\vIdxTwo}}^{N-1}
	\frac{\ulDuration\beta_{\sIdxThree}\ulTotPower{}{\sIdxThree}}{M\rhoD{}{\vIdxTwo}^2\beta_{\vIdxTwo}}
	+
	\sum\limits_{\substack{\sIdxThree=0\\\sIdxThree\neq\vIdxTwo}}^{N-1}
	\sum\limits_{\substack{\sIdxTwo=0\\\sIdxTwo\neq\sIdxThree}}^{N-1}
	\frac{\rhoD{}{\sIdxTwo}^2\beta_{\sIdxThree}\beta_{\sIdxTwo}\ulTotPower{}{\sIdxThree}}{M\rhoP{}{\vIdxTwo}^2\rhoD{}{\vIdxTwo}^2\beta_{\vIdxTwo}^2}
}
\;.
\end{align}
It completes the derivation of \eqref{eqn:sinrNonIterativeFiniteM}.


\section{}
\label{appdx:sinrCalc}
\subsection*{Interference Power of the Iterative Method}
To derive the SINR, using the definition of $ \Delta\mbf{x}_{\vIdxTwo}^{(i)} \triangleq \mbf{x}_\vIdxTwo - \widehat{\mbf{x}}_{\vIdxTwo}^{(i)} $, the channel estimate in \eqref{eqn:iterativeChannelEstimate} can be simplified as
\renewcommand{\vIdxOne}{j}
\renewcommand{\vIdxTwo}{m}
\renewcommand{\sIdxOne}{\ell}
\renewcommand{\sIdxTwo}{k}
\renewcommand{\sIdxThree}{k}
\begin{align}
&\widehat{\mbf{h}}^{(i)}_{\vIdxTwo} 
= 
\mbf{h}_{\vIdxTwo} 
+ 
\frac{1}{\ulDuration\rhoP{}{\vIdxTwo}}
\left(
\!
\sum\limits_{\sIdxThree}
\rhoD{}{\sIdxThree}
\mbf{h}_{\sIdxThree}
\mbf{x}_{\sIdxThree}^T 
-\!\!\!\!
\sum\limits_{\sIdxThree\in\mathcal{U}_{\vIdxOne},\sIdxThree<\vIdxTwo}
\!\!\!
\rhoD{}{\sIdxThree}
\widehat{\mbf{h}}_{\sIdxThree}^{(i)} \lrc{\widehat{\mbf{x}}_{\sIdxThree}^{(i)}}^T \nonumber\right. \\
&\left.
- 
\!\!\!\!
\sum\limits_{\sIdxThree\in\mathcal{U}_{\vIdxOne},\vIdxTwo\leq\sIdxThree\leq N}
\!\!\!\!
\rhoD{}{\sIdxThree}
\widehat{\mbf{h}}_{\sIdxThree}^{(i-1)} \lrc{\widehat{\mbf{x}}_{\sIdxThree}^{(i-1)}}^T+\mbf{W}_\vIdxOne
\right)
\mbf{p}_{\vIdxTwo}^*
\label{eqn:channelEstimateApprox}
\end{align}
where
\renewcommand{\vIdxOne}{j}
\renewcommand{\vIdxTwo}{m}
\renewcommand{\sIdxOne}{\ell}
\renewcommand{\sIdxTwo}{k}
\renewcommand{\sIdxThree}{k}
\begin{align}
&\Delta\mbf{h}_{\vIdxTwo}^{(i)} = -\frac{1}{\ulDuration\rhoP{}{\vIdxTwo}}
\left(	
\sum\limits_{\sIdxThree\in\mathcal{U}_{\vIdxOne},\sIdxThree<\vIdxTwo}
\!\!\!\!
\rhoD{}{\sIdxThree}
\left\{
\mbf{h}_{\sIdxThree}
\lrc{\Delta\mbf{x}_{\sIdxThree}^{(i)} }^T
+ 
\Delta\mbf{h}_{\sIdxThree}^{(i)}
\mbf{x}_{\sIdxThree}^T \right. \right. \nonumber \\
&\left. \left.\!\!-
\Delta\mbf{h}_{\sIdxThree}^{(i)}
\!\!
\lrc{\Delta\mbf{x}_{\sIdxThree}^{(i)}}^T 
\right\}\!
+\!\!\!\!\!
\sum\limits_{\sIdxThree\in\mathcal{U}_{\vIdxOne},\vIdxTwo\leq\sIdxThree\leq N}
\!\!\!\!\!\!\!\!
\rhoD{}{\sIdxThree}
\!
\left\{		
\mbf{h}_{\sIdxThree}
\!
\lrc{\!\Delta\mbf{x}_{\sIdxThree}^{(i-1)}\!}^T\!\!\!\!
\!
+
\!
\Delta\mbf{h}_{\sIdxThree}^{(i-1)}
\mbf{x}_{\sIdxThree}^T	
\right. \right. \nonumber \\
&\left.\left.		
- 
\Delta\mbf{h}_{\sIdxThree}^{(i-1)}
\lrc{\Delta\mbf{x}_{\sIdxThree}^{(i-1)}}^T
\right\}
+
\sum\limits_{\sIdxThree\notin\mathcal{U}_\vIdxOne}
\rhoD{}{\sIdxThree}
\mbf{h}_{\sIdxThree}
\mbf{x}_{\sIdxThree}^T
+
\mbf{W}
\right.\Bigg)
\mbf{p}_{\vIdxTwo}^*\;.
\label{eqn:deltaHdefn}
\end{align}
The received symbols after MF in \eqref{eqn:iterativeDataEstimate} are then given as
\renewcommand{\vIdxOne}{j}
\renewcommand{\vIdxTwo}{m}
\renewcommand{\sIdxOne}{\ell}
\renewcommand{\sIdxTwo}{k}
\renewcommand{\sIdxThree}{k}
\begin{align}
&\widehat{\mbf{x}}_{\vIdxTwo}^T 
= 
\frac{1}{M\rhoD{}{\vIdxTwo}}
\lrc{\mbf{h}_{\vIdxTwo}^H 
	\!\!-\!\!
	\lrc{\Delta\mbf{h}_{\vIdxTwo}^{(i)}}^H}\!\!\!
\left(
\sum\limits_{\sIdxThree=0}^{N-1}\!
\mbf{h}_{\sIdxThree}\!
\lrc{
	\rhoD{}{\sIdxThree}
	\mbf{x}_{\sIdxThree}
	+
	\rhoP{}{\sIdxThree}
	\mbf{p}_\sIdxThree}^T \right. \nonumber \\
&\left. 
+
\mbf{W}
-
\rhoP{}{\vIdxTwo}
\lrc{
	\mbf{h}_{\vIdxTwo}
	-
	\Delta\mbf{h}_{\vIdxTwo}^{(i)}}
\mbf{p}_\vIdxTwo^T
\right)\nonumber \\
&= 
\frac{1}{M}
\|\mbf{h}_{\vIdxTwo}\|^2\mbf{x}_{\vIdxTwo} ^T
+ \sum\limits_{\sIdxThree=1}^{7} \mbf{a}_\sIdxThree^T
\end{align}
where
\begingroup
\allowdisplaybreaks
\begin{align}
\mbf{a}_1 &\triangleq 
\frac{1}{M\rhoD{}{\vIdxTwo}}
\sum\limits_{\substack{\sIdxThree=0\\\sIdxThree\neq \vIdxTwo}}^{N-1}
\mbf{h}_{\vIdxTwo}^H
\mbf{h}_{\sIdxThree}
\lrc{\rhoD{}{\sIdxThree}\mbf{x}_\sIdxThree
	+
	\rhoP{}{\sIdxThree}
	\mbf{p}_\sIdxThree}
\\
\mbf{a}_2 &\triangleq
\frac{1}{M\rhoD{}{\vIdxTwo}}
\lrc{\mbf{h}_{\vIdxTwo}^H\mbf{W}}^T \\
\mbf{a}_3 &\triangleq 
\frac{\rhoP{}{\vIdxTwo}}{M\rhoD{}{\vIdxTwo}}
\mbf{h}_{\vIdxTwo}^H
\Delta\mbf{h}_{\vIdxTwo}^{(i)}\mbf{p}_{\vIdxTwo}\\
\mbf{a}_4 &\triangleq 
-
\frac{1}{M}\lrc{\Delta\mbf{h}_{\vIdxTwo}^{(i)}}^H \mbf{h}_{\vIdxTwo}\mbf{x}_{\vIdxTwo} \\
\mbf{a}_5 &\triangleq -
\frac{1}{M\rhoD{}{\vIdxTwo}}\sum\limits_{\substack{\sIdxThree=0\\\sIdxThree\neq \vIdxTwo}}^{N-1}\lrc{\Delta\mbf{h}_{\vIdxTwo}^{(i)}}^H\mbf{h}_{\sIdxThree}\lrc{\rhoD{}{\sIdxThree}\mbf{x}_{\sIdxThree}+\rhoP{}{\sIdxThree}\mbf{p}_{\sIdxThree}}\\
\mbf{a}_6 &\triangleq
-
\frac{1}{M\rhoD{}{\vIdxTwo}}
\lrc{\lrc{\Delta\mbf{h}_{\vIdxTwo}^{(i)}}^H\mbf{W}}^T\\
\mbf{a}_7 &\triangleq-
\frac{\rhoP{}{\vIdxTwo}}{M\rhoD{}{\vIdxTwo}}\left\|\Delta\mbf{h}_{\vIdxTwo}^{(i)}\right\|^2\mbf{p}_{\vIdxTwo}  \label{eqn:mfSymbolVector} \ .
\end{align}
\endgroup
Under the assumption that the interference power at each of the received symbols is the same, the average interference power of the $ \vIdxTwo\rth $ user at the $ \vIdxOne\rth $ cell is given as
\renewcommand{\vIdxOne}{j}
\renewcommand{\vIdxTwo}{m}
\renewcommand{\sIdxOne}{\ell}
\renewcommand{\sIdxTwo}{k}
\renewcommand{\sIdxThree}{k}
\begin{align}
\spUlInterferencePower_{\vIdxTwo}^{(i)} &= \frac{1}{\ulDuration}\mathbb{E}\left\{\left\|\sum\limits_{\sIdxThree=1}^{7}\mbf{a}_\sIdxThree\right\|^2\right\}
\approx \frac{1}{\ulDuration}
\left[
\mathbb{E}\lrf{\sum\limits_{\sIdxThree=1}^{5}\|\mbf{a}_\sIdxThree\|^2}
\right]
\label{eqn:interferenceApprox}
\end{align}
where the terms $ \mbf{a}_6 $, $ \mbf{a}_7 $, and $ \mbf{a}_p^H\mbf{a}_q, \;\forall p,q $ have been dropped.
Further, it can be shown straightforwardly that
\renewcommand{\vIdxOne}{j}
\renewcommand{\vIdxTwo}{m}
\renewcommand{\sIdxOne}{\ell}
\renewcommand{\sIdxTwo}{k}
\renewcommand{\sIdxThree}{k}
\begin{align}
\mathbb{E}\lrf{\|\mbf{a}_1\|^2} &= 
\frac{\ulDuration}{M\rhoD{}{\vIdxTwo}^2} 
\sum\limits_{\substack{\sIdxThree=0\\\sIdxThree\neq\vIdxTwo}}^{N-1} 
\beta_{\sIdxThree}
\beta_{\vIdxTwo}  
\label{eqn:a1}\\
\mathbb{E}\lrf{\|\mbf{a}_2\|^2} 
&= 
\frac{\ulDuration\sigma^2\beta_{\vIdxTwo}}{M\rhoD{}{\vIdxTwo}^2}  \label{eqn:a2} \ .
\end{align}
Moreover, $  \mathbb{E}\lrf{\|\mbf{a}_3\|^2} $, $  \mathbb{E}\lrf{\|\mbf{a}_4\|^2} $, and $ \mathbb{E}\lrf{\|\mbf{a}_5\|^2} $ can be written as
\renewcommand{\vIdxOne}{j}
\renewcommand{\vIdxTwo}{m}
\renewcommand{\sIdxOne}{\ell}
\renewcommand{\sIdxTwo}{k}
\renewcommand{\sIdxThree}{k}
\begin{align}
&\mathbb{E}\lrf{\|\mbf{a}_3\|^2} 
=
\frac{\rhoP{}{\vIdxTwo}^2}{M^2\rhoD{}{\vIdxTwo}^2}
\mathbb{E}
\left\{
\mbf{h}_{\vIdxTwo}^H
\Delta\mbf{h}_{\vIdxTwo}^{(i)}
\mbf{p}_\vIdxTwo^T 
\mbf{p}_\vIdxTwo^*
\lrc{\Delta\mbf{h}_{\vIdxTwo}^{(i)}}^H
\mbf{h}_{\vIdxTwo}
\right\} \nonumber \\
&
=
\frac{\ulDuration\rhoP{}{\vIdxTwo}^2}{M^2\rhoD{}{\vIdxTwo}^2}\mathbb{E}\lrf{\lrc{\Delta\mbf{h}_{\vIdxTwo}^{(i)}}^H\mbf{h}_{\vIdxTwo}\mbf{h}_{\vIdxTwo}^H\Delta\mbf{h}_{\vIdxTwo}^{(i)}}\label{eqn:a3}\\
&\mathbb{E}\lrf{\|\mbf{a}_4\|^2} 
= 
\frac{1}{M^2}
\mathbb{E}
\left\{
\lrc{\Delta\mbf{h}_{\vIdxTwo}^{(i)}}^H 
\mbf{h}_{\vIdxTwo}
\mbf{x}_{\vIdxTwo}^T
\mbf{x}_{\vIdxTwo}^*
\mbf{h}_{\vIdxTwo}^H
\Delta\mbf{h}_{\vIdxTwo}^{(i)}
\right\} \nonumber\\
&=
\frac{1}{M^2}
\mathbb{E}\lrf{\mbf{x}_{\vIdxTwo}^T\mbf{x}_\vIdxTwo^*}
\mathbb{E}\lrf{\lrc{
		\Delta\mbf{h}_{\vIdxTwo}^{(i)}}^H 
	\mbf{h}_{\vIdxTwo}
	\mbf{h}_{\vIdxTwo}^H
	\Delta\mbf{h}_{\vIdxTwo}^{(i)}}\nonumber\\
&=\frac{\ulDuration}{M^2}
\mathbb{E}\lrf{\lrc{
		\Delta\mbf{h}_{\vIdxTwo}^{(i)}}^H 
	\mbf{h}_{\vIdxTwo}
	\mbf{h}_{\vIdxTwo}^H
	\Delta\mbf{h}_{\vIdxTwo}^{(i)}} \label{eqn:a4}
\end{align}
and
\begin{align}
&\mathbb{E}\lrf{\|\mbf{a}_5\|^2}
= 
\frac{1}{M^2\rhoD{}{\vIdxTwo}^2} 
\sum\limits_{\substack{\sIdxOne=0\\\sIdxOne\neq \vIdxTwo}}^{N-1}
\sum\limits_{\substack{\sIdxTwo=0\\\sIdxTwo\neq\vIdxTwo}}^{N-1}
\mathbb{E}
\lrf{
	\lrc{\Delta\mbf{h}_{\vIdxTwo}^{(i)}}^H	
	\mbf{h}_{\sIdxOne}
	\mbf{h}_{\sIdxTwo}^H
	\Delta\mbf{h}_{\vIdxTwo}^{(i)}
} 		
\nonumber\\
&\qquad\qquad\qquad
\times
\mathbb{E}
\lrf{	
	\lrc{\rhoD{}{\sIdxOne}\mbf{x}_\sIdxOne+\rhoP{}{\sIdxOne}\mbf{p}_\sIdxOne}^H
	\lrc{\rhoD{}{\sIdxTwo}\mbf{x}_\sIdxTwo+\rhoP{}{\sIdxTwo}\mbf{p}_\sIdxTwo}	
}
\nonumber\\
&=
\frac{\ulDuration}{M^2\rhoD{}{\vIdxTwo}^2} 
\mathbb{E}
\lrf{	
	\lrc{\Delta\mbf{h}_{\vIdxTwo}^{(i)}}^H
	\lrc{\sum\limits_{\substack{\sIdxTwo=0\\\sIdxTwo\neq \vIdxTwo}}^{N-1}
		\mbf{h}_{\sIdxTwo}\mbf{h}_{\sIdxTwo}^H}
	\Delta\mbf{h}_{\vIdxTwo}^{(i)}}\label{eqn:a5} \ .
\end{align}
Summing up \eqref{eqn:a3}, \eqref{eqn:a4}, and \eqref{eqn:a5}, we obtain
\renewcommand{\vIdxOne}{j}
\renewcommand{\vIdxTwo}{m}
\renewcommand{\sIdxOne}{\ell}
\renewcommand{\sIdxTwo}{k}
\renewcommand{\sIdxThree}{k}
\begin{align}
&\mathbb{E}\lrf{\sum\limits_{\sIdxThree=3}^{5}\|\mbf{a}_\sIdxThree\|^2} 
=
\frac{\ulDuration}{M^2\rhoD{}{\vIdxTwo}^2} 
\nonumber\\
&\qquad\times
\mathbb{E}
\left\{
\lrc{\Delta\mbf{h}_{\vIdxTwo}^{(i)}}^H 	
\lrc{
	\sum\limits_{\sIdxTwo=0}^{N-1}
	\mbf{h}_{\sIdxTwo}
	\mbf{h}_{\sIdxTwo}^H}
\Delta\mbf{h}_{\vIdxTwo}^{(i)}
\right\} \label{eqn:sumA4A5} \ .
\end{align}
Now, let $ \psi_{\vIdxTwo}^{(i)} $ be defined as the second term in \eqref{eqn:sumA4A5}, i.e., 
\renewcommand{\vIdxOne}{j}
\renewcommand{\vIdxTwo}{m}
\renewcommand{\sIdxOne}{\ell}
\renewcommand{\sIdxTwo}{n}
\renewcommand{\sIdxThree}{k}
\begin{align}
&\psi_{\vIdxTwo}^{(i)}\big|_{i\geq 1}\triangleq\mathbb{E}\lrf{
	\lrc{\Delta\mbf{h}_{\vIdxTwo}^{(i)}}^H
	\lrc{
		\sum\limits_{\sIdxTwo=0}^{N-1}
		\mbf{h}_{\sIdxTwo}
		\mbf{h}_{\sIdxTwo}^H
	}
	\Delta\mbf{h}_{\vIdxTwo}^{(i)}
}\;.  \label{eqn:sinrSumTermInterimstep}
\end{align}
Using \eqref{eqn:deltaHdefn} and the simplifications \ref{S1} to \ref{S4}, \eqref{eqn:sinrSumTermInterimstep} can be simplified to obtain \eqref{eqn:psiDefn}.
Substituting \eqref{eqn:a1}, \eqref{eqn:a2}, \eqref{eqn:sumA4A5}, and \eqref{eqn:psiDefn} into \eqref{eqn:interferenceApprox}, $ \spUlInterferencePower_{\vIdxTwo}^{(i)} $ can be obtained as
\renewcommand{\vIdxOne}{j}
\renewcommand{\vIdxTwo}{m}
\renewcommand{\sIdxOne}{\ell}
\renewcommand{\sIdxTwo}{n}
\renewcommand{\sIdxThree}{k}
\begin{align}
\spUlInterferencePower_{\vIdxTwo}^{(i)} 
&
\approx 
\frac{1}{M\rhoD{}{\vIdxTwo}^2} 
\sum\limits_{\substack{\sIdxThree=0\\\sIdxThree\neq \vIdxTwo}}^{N-1}
\beta_{\sIdxThree}
\beta_{\vIdxTwo} 
+
\frac{\sigma^2\beta_{\vIdxTwo}}{M\rhoD{}{\vIdxTwo}^2}
+
\frac{1}{M^2\rhoD{}{\vIdxTwo}^2} 
\psi_{\vIdxTwo}^{(i)}\;.
\label{eqn:appdxInterferenceExpression}
\end{align}
It completes the derivation of \eqref{eqn:interferenceExpression}.



\section{}
\label{appdx:iterativeChannelEstimationThreshold}

\subsection*{Choice of the Set of Users $ \mathcal{U}_{\vIdxTwo}^{(i)} $}
Let $ \mathcal{S} $ be a set of the $ KL $ users in the system and let $ \powerset{\mathcal{S}} $ be its power set. In addition, for the sake of clarity, let the additional argument $ \mathcal{U}_{\vIdxTwo}^{(i)} $ be added to the functions $ \spUlInterferencePower_{\vIdxTwo}^{(i)} $ and $ \psi_{\vIdxTwo}^{(i)} $ in this section.
Now, the optimal set $ \mathcal{U}_{\vIdxTwo}^{(i)} $ can be obtained by solving the following optimization problem
\begin{equation}
\mathcal{U}_{\vIdxTwo}^{(i)} = \arg\min_{\mathcal{U}\in\powerset{\mathcal{S}}} \lrf{\spUlInterferencePower_{\vIdxTwo}^{(i)}\lrc{\mathcal{U}}}\;.
\label{eqn:thresholdOptimizationProblem}
\end{equation}
Substituting \eqref{eqn:interferenceExpression} into \eqref{eqn:thresholdOptimizationProblem} yields 
\begin{equation}
\mathcal{U}_{\vIdxTwo}^{(i)} = \arg\min_{\mathcal{U}\in\powerset{\mathcal{S}}} \lrf{\psi_{\vIdxTwo}^{(i)}\lrc{\mathcal{U}}}\;.
\label{eqn:thresholdOptimizationProblemPsiVersion}
\end{equation}
Now, $ \psi_{\vIdxTwo}^{(i)}(\mathcal{U}) $ can be rewritten as
\begin{align}
\psi_{\vIdxTwo}^{(i)}(\mathcal{U})
&= 
c
+
\sum\limits_{\sIdxTwo=0}^{N-1} 
\left\{
\xi_{\sIdxTwo}
\indicator{\sIdxTwo\notin\mathcal{U}}
+
\epsilon_{\sIdxTwo}^{(i)}(\mathcal{U}) 
\indicator{\sIdxTwo\in\mathcal{U},\sIdxTwo<\vIdxTwo} \nonumber \right. \\
&\quad\left.
+
\epsilon_{\sIdxTwo}^{(i-1)}(\mathcal{U})
\indicator{\sIdxTwo\in\mathcal{U},\sIdxTwo\geq\vIdxTwo}
\right\} \label{eqn:psiRedefinition}
\end{align}
where $ c $,  $ \xi_{\sIdxTwo} $, and $ \epsilon_{\sIdxTwo}^{(i)}(\mathcal{U}) $ are defined as
\begin{align}
c 
&\triangleq 
\frac{M\sigma^2}{\ulDuration\rhoP{}{\vIdxTwo}^2}
\lrc{
	\sum\limits_{\sIdxThree=0}^{N-1}
	\beta_{\sIdxThree}	
} \\
\xi_{\sIdxTwo}
&\triangleq
\frac{M^2\rhoD{}{\sIdxTwo}^2}{\ulDuration\rhoP{}{\vIdxTwo}^2}
\left\{			
\beta_{\sIdxTwo}^2
+
\frac{1}{M}		
\sum\limits_{\sIdxThree=0}^{N-1}
\beta_{\sIdxThree}
\beta_{\sIdxTwo}
\right\}\label{eqn:xiDefn}
\\
\epsilon_{\sIdxTwo}^{(i)}(\mathcal{U})
&\triangleq
\frac{M^2\rhoD{}{\sIdxTwo}^2}{\ulDuration\rhoP{}{\vIdxTwo}^2}
\left\{		
\beta_{\sIdxTwo}^2
\alpha_{\sIdxTwo}^{(i)}
+
\sum\limits_{\sIdxThree=0}^{N-1}
\frac{1}{M}
\beta_{\sIdxThree}
\beta_{\sIdxTwo}
\alpha_{\sIdxTwo}^{(i)} \right. \nonumber \\
&\quad\left.
+
\frac{\lrc{1+\alpha_{\sIdxTwo}^{(i)}}}{M^2}
\psi_{\sIdxTwo}^{(i)}(\mathcal{U})
\right\}\;.
\label{eqn:epsilonDefn}
\end{align}
It can be seen from \eqref{eqn:psiRedefinition} that the optimization problem \eqref{eqn:thresholdOptimizationProblemPsiVersion} is separable over the user indices, implying that the decision to include user $ \sIdxTwo $ in $ \mathcal{U}_{\vIdxTwo}^{(i)} $ is independent of the other $ N-1 $ users. Therefore, the channel and data estimates of user $ \sIdxTwo $ are used in the $ i\rth $ iteration if the following condition is satisfied
\begin{align}
\sIdxTwo\in\mathcal{U}_{\vIdxTwo}^{(i)}
\;\;\mathrm{iff}\;\; 
\psi_{\vIdxTwo}^{(i)}\big|_{\sIdxTwo\in\mathcal{U}_{\vIdxTwo}^{(i)}} < 
\psi_{\vIdxTwo}^{(i)} \big|_{\sIdxTwo\notin\mathcal{U}_{\vIdxTwo}^{(i)}}\;.
\label{eqn:optimalThresholdCondition} 
\end{align}
From \eqref{eqn:psiRedefinition} and \eqref{eqn:optimalThresholdCondition}, the set $ \mathcal{U}_{\vIdxTwo}^{(i)} $ is obtained as
\begin{align}
\mathcal{U}_{\vIdxTwo}^{(i)} &= 
\left\{
\sIdxTwo\in\mathbb{N}\;\vrule\;
\epsilon_{\sIdxTwo}^{(i)}(\mathcal{U}) < \xi_{\sIdxTwo}\;\mathrm{when}\;\sIdxTwo<\vIdxTwo\right.
\nonumber\\
&\qquad
\left.
\;\mathrm{and}\;\epsilon_{\sIdxTwo}^{(i-1)}(\mathcal{U}) < \xi_{\sIdxTwo}\;\mathrm{when}\;\sIdxTwo\geq\vIdxTwo
\right\} \;.
\end{align}
Equivalently, using \eqref{eqn:xiDefn} and \eqref{eqn:epsilonDefn}, the above expression simplifies to
\begin{align}
\mathcal{U}_{\vIdxTwo}^{(i)} &=
\left\{
\sIdxTwo\in\mathbb{N}\;\vrule\;
\alpha_{\sIdxTwo}^{(i)} < \gamma_{\sIdxTwo}^{(i)} \;\mathrm{when}\; \sIdxTwo<\vIdxTwo \right. \nonumber \\
&\qquad
\left.
\mathrm{and}\;
\alpha_{\sIdxTwo}^{(i-1)} < \gamma_{\sIdxTwo}^{(i-1)} \;\mathrm{when}\; \sIdxTwo\geq\vIdxTwo
\right\}
\label{eqn:optimalThresholdAlphaEqn}
\end{align}
where 
\begin{equation}
\gamma_{\sIdxTwo}^{(i)} \triangleq \frac{\lrf{	
		\beta_{\sIdxTwo}^2
		+
		\frac{1}{M}
		\sum\limits_{\sIdxThree=0}^{N-1}
		\beta_{\sIdxTwo}
		\beta_{\sIdxThree}
		-
		\frac{\psi_{\sIdxTwo}^{(i)}\big|_{\sIdxTwo\in\mathcal{U}_{\vIdxTwo}^{(i)}}}{M^2}
	}}
	{
		\lrf{	
			\beta_{\sIdxTwo}^2
			+
			\frac{1}{M}		
			\sum\limits_{\sIdxThree=0}^{N-1}
			\beta_{\sIdxTwo}
			\beta_{\sIdxThree}				
			+
			\frac{\psi_{\sIdxTwo}^{(i)}\big|_{\sIdxTwo\in\mathcal{U}_{\vIdxTwo}^{(i)}}}{M^2}
		}
	} \;.
	\label{eqn:gammaDefn}
	\end{equation}
	If $ \mbf{x}_\vIdxTwo $ takes values from the $ P $-QAM constellation, then substituting \eqref{eqn:alphaDefnPQAM} into \eqref{eqn:optimalThresholdAlphaEqn}, the set $ \mathcal{U}_{\vIdxTwo}^{(i)} $ can be obtained as
	\begin{align}
	\mathcal{U}_{\vIdxTwo}^{(i)} &=
	\left\{
	\sIdxTwo\in\mathbb{N}
	\;\vrule\;
	\spUlInterferencePower_{\sIdxTwo}^{(i)}
	< 
	f_{\sIdxTwo}^{(i)}
	\;\mathrm{when}\;
	\sIdxTwo<\vIdxTwo	
	\right.\nonumber
	\\
	&\qquad
	\mathrm{and}\;\;
	\left.
	\spUlInterferencePower_{\sIdxTwo}^{(i)}
	<
	f_{\sIdxTwo}^{(i-1)}
	\;\mathrm{when}\;
	\sIdxTwo\geq\vIdxTwo
	\right\}
	\label{eqn:optimalThresholdConditionQAM}
	\end{align}
	where $ f_{\sIdxTwo}^{(i)} $ is defined as
	\begin{align}
	f_{\sIdxTwo}^{(i)}&\triangleq \frac{3}{P-1} Q^{2}
	\lrc{
		\frac{\sqrt{P}\lrc{\sqrt{P}+1}\gamma_{\sIdxTwo}^{(i)}}{24}			
	}\;.
	\end{align}
	However, since the decision rules are based on approximate SINR expressions, it is worth commenting that the reliability of the decision rule in \eqref{eqn:optimalThresholdAlphaEqn} decreases with increasing user and iteration indices. Alternatively, a fixed and conservative decision rule can be used to obtain $ \mathcal{U} $ as follows
	\begin{align}
	\mathcal{U}_{\mathrm{fixed}} &= 
	\left\{
	\vIdxTwo\in\mathbb{N}
	\;\vrule\;
	\spUlInterferencePower_{\vIdxTwo}^{(2)}\lrc{\lrf{\vIdxTwo}} < \spUlInterferencePower_{\vIdxTwo}^{(2)}\lrc{\varnothing} = \spUlInterferencePower_{\vIdxTwo}^{(1)}\lrc{\varnothing}
	\right\} \;.
	\label{eqn:suboptimalThresholdCondition}
	\end{align}
	The decision rule in \eqref{eqn:suboptimalThresholdCondition} results in a set $ \mathcal{U}_{\mathrm{fixed}} $ that is computed at the beginning of the first iteration and is left unchanged for the subsequent iterations. 
	

\section{}
\label{appdx:alphaPQAMderivation}
\subsection*{Derivation of $ \alpha_{\vIdxTwo}^{(i)} $ for a $ P $-QAM constellation}
For $ P $-QAM constellation and $ i\geq 1 $, the integral over $ \mbf{x}_{\vIdxTwo} $ in \eqref{eqn:alphaDefn} reduces to a summation, which can be written as
\begin{align}
\alpha_{\vIdxTwo}^{(i)} 
&=
\sum\limits_{x\in\chi}\int
\left|
x
-
\eta\lrc{
	x 
	- 
	e
}\right|^2
p_{\xTildeError{\vIdxTwo}^{(i)}}
\lrc{
	e
}
p_{\mbf{x}_\vIdxTwo}
\lrc{x}
de\ .
\end{align}
Since the $ P $ symbols are equally likely, $ p_{\mbf{x}_\vIdxTwo}(x) = 1/P, \; \forall\; x $ and under the assumption that the errors $ x-\eta\lrc{x-e} $ are dominated by the closest neighboring symbols, the above equation reduces to
\begin{align}
\alpha_{\vIdxTwo}^{(i)} 
&= \frac{1}{P}
\sum\limits_{x\in\chi}
d_x^2 
k_x
Q\lrc{\frac{\frac{d_x}{2}}{\sqrt{\frac{\spUlInterferencePower_{\vIdxTwo}^{(i)}}{2}}}}
\label{eqn:alphaDerivationInterimStep}
\end{align}
where $ d_x $ is the distance between the symbol $ x $ and its closest neighbor and $ k_x $ is the number of symbols at a distance of $ d_x $ from $ x $. The Q-function in the above equation results from the assumption on the statistics of $ \xTildeError{\vIdxTwo}^{(i)} $. For a unit-power $ P $-QAM constellation, $ d_x = \sqrt{6/P-1}, \;\forall x$ \cite{proakis2008digital}. In addition, it can be easily verified that $ k_x = 2 $ for the $ 4 $ corner symbols, $ k_x = 3 $ for the $ (\sqrt{P}-2)4 $ symbols on the outer edges, and $ k_x = 4 $ for the remaining $ P-4\sqrt{P}+4 $ symbols. Substituting these values into \eqref{eqn:alphaDerivationInterimStep} yields
\begin{align}
\alpha_{\vIdxTwo}^{(i)}\big|_{i\geq 1} = \frac{24}{\sqrt{P}\lrc{\sqrt{P}+1}}Q\lrc{\sqrt{\frac{\frac{3}{(P-1)}}{\spUlInterferencePower_{\vIdxTwo}^{(i)}}}} \;.
\end{align}
Moreover, since $ \Delta\mbf{x}_{\sIdxOne,\vIdxTwo}^{(0)} = \mbf{x}_{\sIdxOne,\vIdxTwo} $, the value of $ \alpha_{\vIdxTwo}^{(0)} $ is $ 1 $. It completes the derivation of \eqref{eqn:alphaDefnPQAM}.

\bibliographystyle{IEEEtran}
\bibliography{IEEEabrv,paperBibliography.bib}

\vspace{-0.7cm}
\begin{IEEEbiography}[{\includegraphics[width=1in,height=1.25in,clip,keepaspectratio]{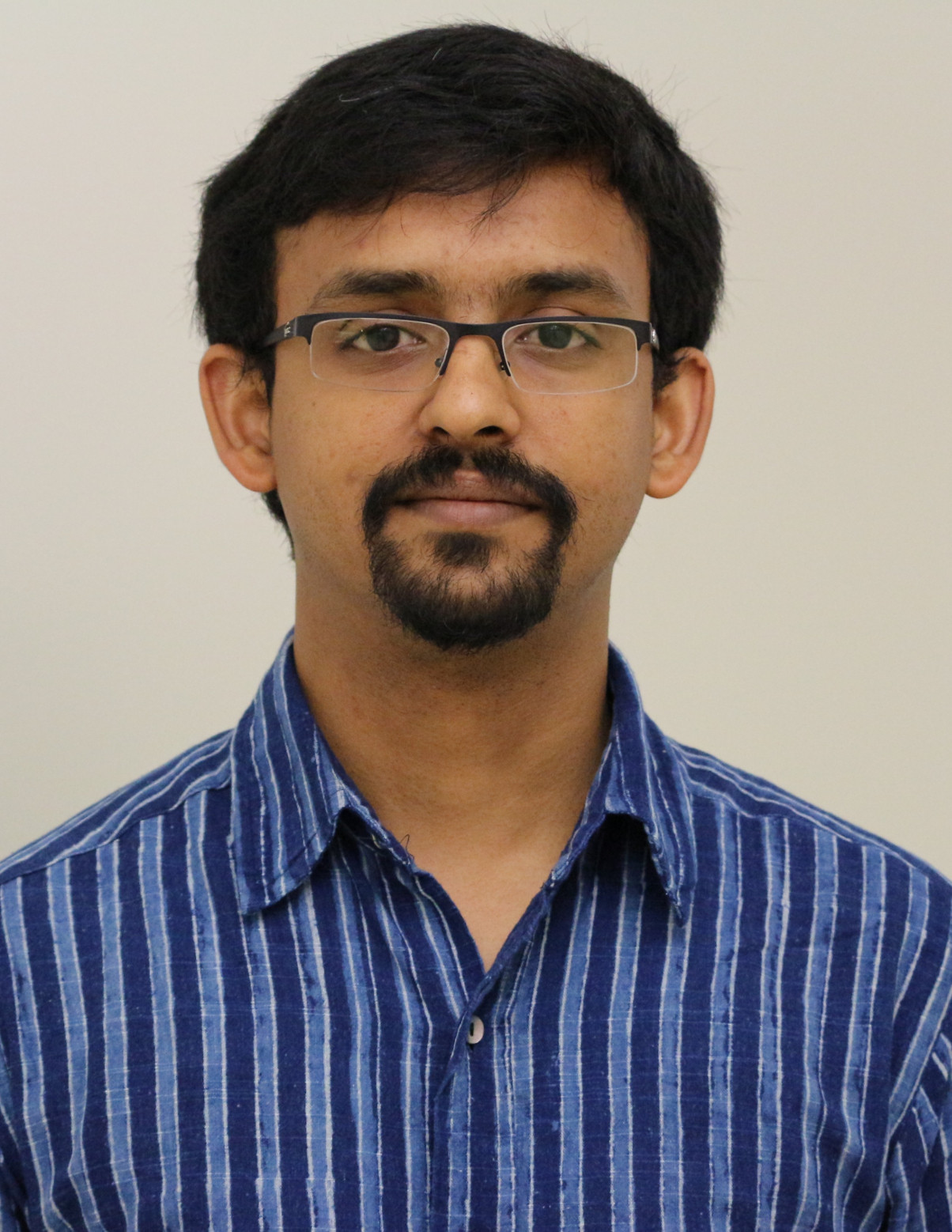}}]{Karthik Upadhya}
	received the B.E. degree in electronics and communication engineering from Visvesvaraya Technological University, India in 2007 and the M.Tech degree in communication systems from Indian Institute of Technology Madras, India in 2011. He is currently a Ph.D. student at Aalto University, Finland. His research interests include wireless communications and array signal processing. Before joining the Ph.D. program, he worked as a Member of Technical Staff at Saankhya Labs, India from 2011 to 2013 and as a research assistant at the Indian Institute of Science (IISc), India from 2013 to 2014.
\end{IEEEbiography}
\vspace{-0.5cm}

\begin{IEEEbiography}[{\includegraphics[width=1in,height=1.25in,clip,keepaspectratio]{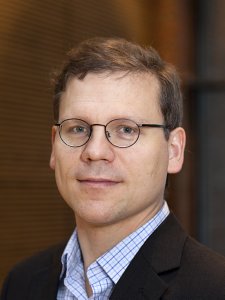}}]{Sergiy Vorobyov}
	(M'02-SM'05) received the M.Sc. and Ph.D. degrees in systems and control from Kharkiv National University of Radio Electronics, Ukraine, in 1994 and 1997, respectively.
	He is a Professor with the Department of Signal Processing and Acoustics, Aalto University, Finland. He has been previously with the University of Alberta, Alberta, Canada as an Assistant Professor from 2006 to 2010, Associate Professor from 2010 to 2012, and became Full Professor there in 2012. Since his graduation, he also held various research and faculty positions at Kharkiv National University of Radio Electronics, Ukraine; the Institute of Physical and Chemical Research (RIKEN), Japan; McMaster University, Canada; Duisburg-Essen University and Darmstadt University of Technology, Germany; and the Joint Research Institute between Heriot-Watt University and Edinburgh University, U.K. His research interests include optimization and liner algebra methods in signal processing and communications; statistical and array signal processing; sparse signal processing; estimation and detection theory; sampling theory; and multi-antenna, very large, cooperative, and cognitive systems.
	
		Dr. Vorobyov is a recipient of the 2004 IEEE Signal Processing Society Best Paper Award, the 2007 Alberta Ingenuity New Faculty Award, the 2011 Carl Zeiss Award (Germany), the 2012 NSERC Discovery Accelerator Award, and other awards. He is serving as Area Editor for IEEE Signal Processing Letters since 2016. He served as an Associate Editor for the IEEE Transactions on Signal Processing from 2006 to 2010 and for the IEEE Transactions on Signal Processing Letters from 2007 to 2009. He was a member of the Sensor Array and Multi-Channel Signal Processing and Signal Processing for Communications and Networking Technical Committees of the IEEE Signal Processing Society from 2007 to 2012 and from 2010 to 2016, respectively. He has served as the Track Chair for Asilomar 2011, Pacific Grove, CA, the Technical Co-Chair for IEEE CAMSAP 2011, Puerto Rico, and the Tutorial Chair for ISWCS 2013, Ilmenau, Germany.
\end{IEEEbiography}
\vspace{-0.55cm}
\begin{IEEEbiography}[{\includegraphics[width=1in,height=1.25in,clip,keepaspectratio]{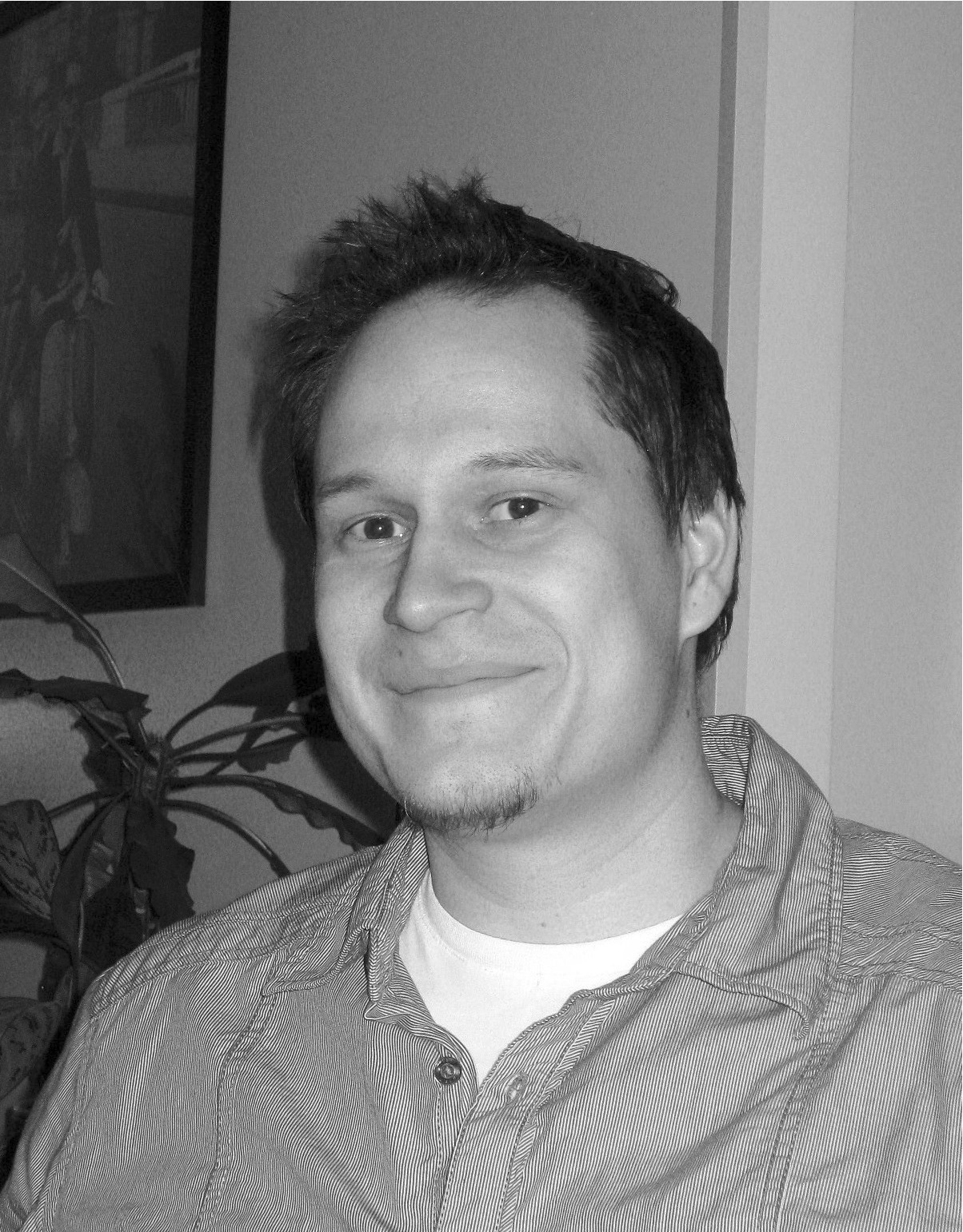}}]{Mikko Vehkapera} (M'10) received the Ph.D.\ degree from Norwegian University of Science and Technology (NTNU), Trondheim, Norway, in 2010. Between 2010--2013 he was a post-doctoral researcher at School of Electrical Engineering, and the ACCESS Linnaeus Center, KTH Royal Institute of Technology, Sweden, and 2013--2015 an Academy of Finland Postdoctoral Researcher at Aalto University School of Electrical Engineering, Finland. He is now an assistant professor (lecturer) at University of Sheffield, Department of  Electronic and Electrical Engineering, United Kingdom. He held visiting appointments at Massachusetts Institute of Technology (MIT), US, Kyoto University and Tokyo Institute of Technology, Japan, and University of Erlangen-Nuremberg, Germany.  His research interests are in the field of wireless communications, information theory and signal processing. 
	Dr.\ Vehkapera was a co-recipient for the Best Student Paper Award at IEEE International Conference on Networks (ICON2011) and IEEE Sweden Joint VT-COM-IT Chapter Best Student Conference Paper Award 2015.
\end{IEEEbiography}

\end{document}